\documentclass[10pt, twocolumn, letterpaper]{article}

\PassOptionsToPackage{numbers}{natbib}

\usepackage[nonatbib, preprint]{cig}

\usepackage[utf8]{inputenc}
\usepackage[T1]{fontenc}

\usepackage{amsmath, amssymb, amsfonts, mathtools, bm}
\usepackage{nicefrac}
\usepackage{bbm}
\usepackage{siunitx}
\usepackage{pifont}
\usepackage{algorithm}
\usepackage{algpseudocode}
\usepackage{amsthm}
\usepackage{calc}
\usepackage{cite}
\usepackage{enumitem}
\usepackage[table]{xcolor}
\definecolor{darkblue}{rgb}{0, 0, 0.85}
\definecolor{lightgreen}{rgb}{.85,1,.85}
\definecolor{lightred}{rgb}{1,.85,.85}
\definecolor{lightblue}{rgb}{.85,.85,1}
\definecolor{pink}{HTML}{EB346F}
\definecolor{cvprblue}{rgb}{0.21,0.49,0.74}

\usepackage{hyperref}
\hypersetup{
    colorlinks = true,
    citecolor = cvprblue,
    linkcolor = black
}

\usepackage{graphicx}
\usepackage{tabularx, threeparttable, tabularray}
\UseTblrLibrary{booktabs}
\usepackage{booktabs}
\usepackage{multirow}
\usepackage{wrapfig}
\usepackage{arydshln}
\usepackage{hhline}
\usepackage{caption}
\usepackage{diagbox}

\usepackage{microtype}
\usepackage{setspace}
\usepackage{lipsum}
\usepackage{soul}

\usepackage{tikz}
\usepackage{outlines} 

\theoremstyle{plain}

\newcommand{\hlgreen}[1]{{\sethlcolor{lightgreen}\hl{#1}}}

\newcommand{\hlblue}[1]{{\sethlcolor{lightblue}\hl{#1}}}

\DeclarePairedDelimiterX{\infdivx}[2]{(}{)}{#1\;\delimsize\|\;#2}

\renewcommand{\vec}[1]{\bm{#1}}

\newcommand{\X}{{\vec{X}}}  
  
\newcommand{\Z}{{\vec{Z}}}  
\newcommand{\Q}{{\vec{Q}}}

\def\x{\vec{x}}  
\def\r{\vec{r}}  
  
\def\z{\vec{z}}

\def\P{\mathbb{P}}
\def\Q{\mathbb{Q}}

\usepackage{pifont}
\newcommand{\cmark}{\textcolor{green!60!black}{\ding{51}}} 
\newcommand{\xmark}{\textcolor{red}{\ding{55}}}            

\newcolumntype{x}[1]{>{\centering\arraybackslash}p{#1pt}}
\newcolumntype{y}[1]{>{\raggedright\arraybackslash}p{#1pt}}
\newcolumntype{z}[1]{>{\raggedleft\arraybackslash}p{#1pt}}

\definecolor{selectcolor}{gray}{.85}

\renewcommand{\paragraph}[1]{\vspace{1.25mm}\noindent\textbf{#1}}

\newcommand{\app}{\raise.17ex\hbox{$\scriptstyle\sim$}}

\definecolor{darkblue}{rgb}{0, 0, 0.5}
\definecolor{iccvblue}{rgb}{0.21,0.49,0.74}
\definecolor{darkgreen}{rgb}{0.16,0.85,0.43}
\definecolor{lightyellow}{RGB}{240,230,149}
\definecolor{darkred}{RGB}{220,20,60}

\title{Next-Dense-Stride Prediction for \\ Multimodal Autoregressive Visual Modeling}

\vspace{5em}
\author{%
\normalsize Chicago Y. Park$^{1,}$\thanks{This work was done during an internship at GE HealthCare.} \quad Jialin Mao$^2$ \quad Xiaojian Xu$^2$ \\[0.7em] Taha Kass-Hout$^2$ \quad Ulugbek S. Kamilov$^1$ \quad Cao Xiao$^2$ \\[0.7em]
\small $^1$\textnormal{University of Wisconsin--Madison} \quad $^2$\textnormal{GE HealthCare}\\ [0.7em]
\footnotesize \texttt{\{chicago.park, kamilov\}@wisc.edu} \\
\footnotesize \texttt{\{jialin.mao, xiaojian.xu, taha.kass-hout, cao.xiao\}@gehealthcare.com} \\
}

\usepackage{geometry}
\usepackage{xcolor}
\usepackage{listings}
\usepackage{algorithm}
\usepackage{caption}
\usepackage[table]{xcolor}

\definecolor{codePink}{HTML}{D63384}  
\definecolor{codeTeal}{HTML}{458588}  
\definecolor{codeBlue}{HTML}{2A52BE}  
\definecolor{codeBlack}{HTML}{000000} 
\definecolor{maskpink}{RGB}{255,0,255}    
\definecolor{maskgreen}{RGB}{0,200,0}     

\usepackage{booktabs}
\usepackage[table]{xcolor}
\usepackage{array}
\usepackage{graphicx}
\usepackage{array}

\usepackage{booktabs}
\usepackage{xcolor}
\usepackage{pifont}
\usepackage{array}
\usepackage[dvipsnames]{xcolor}
\newcommand{\ourmethod}{DenseAR}

\begin{document}

\maketitle

\begin{abstract}
We introduce \ourmethod, a new generative paradigm that reformulates autoregressive image generation as coarse-to-fine next-dense-stride prediction using a compact single-scale tokenizer.
Our key insight is that traversing a single-scale latent grid with progressively denser strides naturally captures the transition from global structure to fine detail.
This addresses two limitations of existing autoregressive models at once: the slow inference of raster-order autoregression, which \ourmethod\ avoids by predicting multiple tokens in parallel, and the heavy cost of multi-scale approaches, which need long, multi-resolution token sequences to achieve coarse-to-fine prediction. 
Building on our efficient framework and the flexibility of autoregressive modeling, we further extend DenseAR to a unified model that handles multiple modalities and imaging tasks within a single backbone.
We validate \ourmethod\ on both medical and natural images.
On multi-contrast brain MRI, a single \ourmethod\ model unifies cross-modal translation, modality-conditioned generation, and tumor segmentation, while remaining competitive with task-specific methods.
On ImageNet, \ourmethod\ improves class-conditional generation quality (FID and IS) over both a single-grid baseline without stride ordering and a multi-scale tokenizer-based baseline.
Code is available \href{https://github.com/uw-cig/DenseAR}{\textcolor{Plum}{\texttt{https://github.com/uw-cig/DenseAR}}}.

\end{abstract}

\section{Introduction}
\label{sec:intro}
Visual generative modeling has advanced rapidly, driven in large part by diffusion models, which produce detailed, coherent images through iterative denoising~\cite{ddpm, scoresde, park2024randomwalks, park2026deepparameterinterpolation}. Following the success of Large Language Models (LLMs), an alternative paradigm treats images as discrete sequences: a visual tokenizer compresses an image into vector-quantized (VQ) tokens, and an autoregressive (AR) model predicts these tokens one by one, like words in a sentence~\cite{llamagen, var, rar}.
Within this paradigm, how visual tokens are ordered for generation has emerged as a central design axis that trades quality against efficiency: two orders dominate visual AR---raster order, which generates tokens row by row, and next-scale order, used by visual autoregressive (VAR) modeling~\cite{var}, which generates a sequence of token maps from low to high resolution.
By generating global structure before fine detail---mirroring how images are naturally composed---this coarse-to-fine order substantially improves generation quality over raster-order AR~\cite{var}.

\begin{figure}[t]
\begin{center}
\includegraphics[width=.45\textwidth]{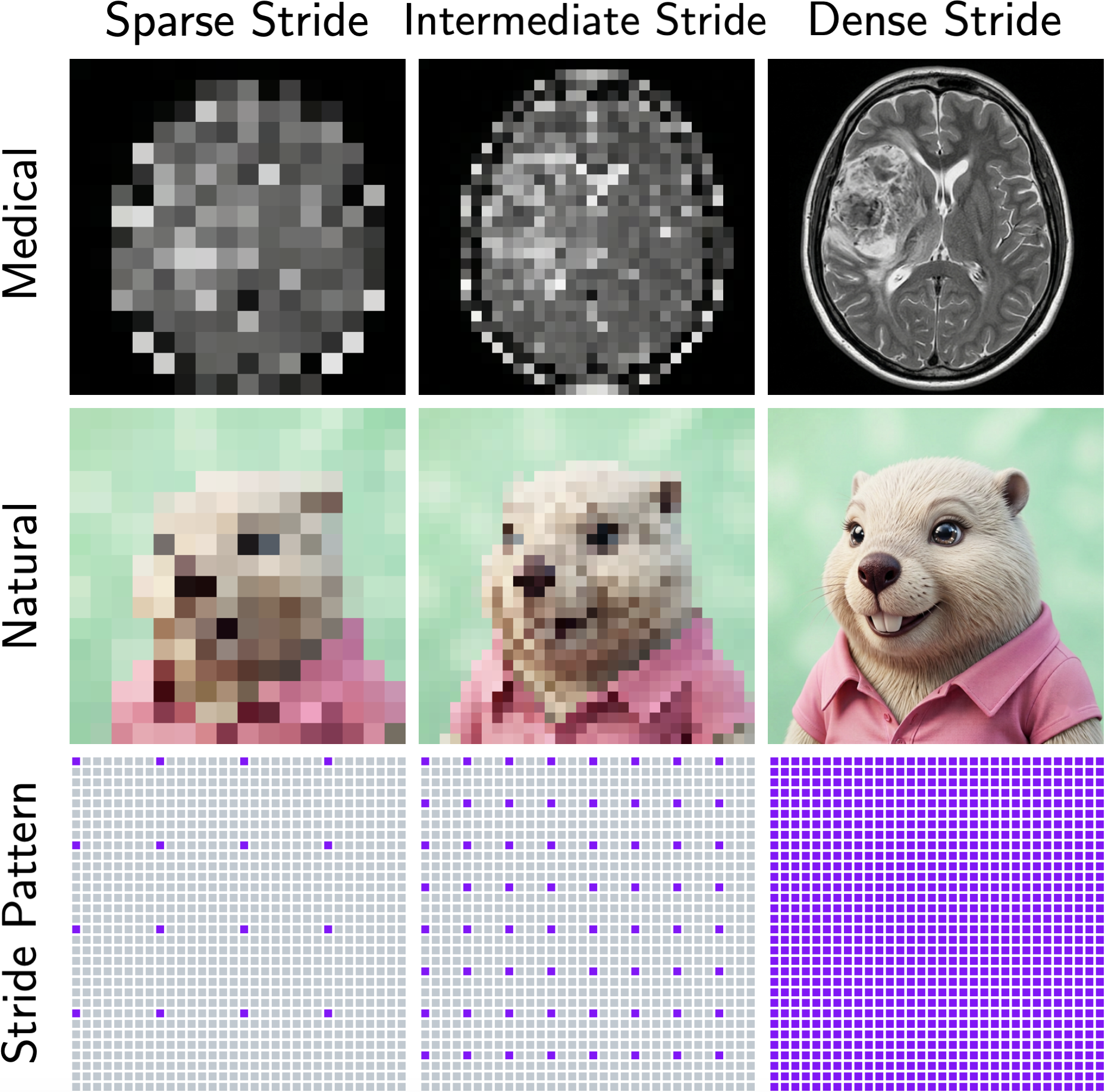}
\end{center}
\caption{Intuition for how stride density controls coarse-to-fine generation: sparse strides capture global structure, while denser strides add fine detail, across medical and natural images.
The bottom row visualizes the \textcolor{Plum}{sampling density} at each stride level.
This figure is illustrative only: we stride in pixel space here for intuition, whereas \ourmethod\ strides on the latent token grid, never on raw pixels.}
\vspace{-.3cm}
\label{fig:stride_intuition}
\end{figure}

Each order has a complementary strength and weakness.
Raster order~\cite{llamagen, Esseretal2021VQGAN} operates on a single, compact latent grid, but generates tokens strictly one after another; this sequential decoding makes inference slow.
Next-scale order~\cite{var, Infinity} instead achieves coarse-to-fine generation, but at a cost: it represents each image as a stack of token maps at several resolutions.
The model processes all multi-scale tokens at once, so the sequence becomes far longer than a single grid, sharply raising memory and compute.
This worsens in conditional generation like image translation and editing, where each source image adds its own multi-scale stack on top of the target~\cite{liu2025cyclevar, mao2025varedit}.
AR is therefore caught between the slow generation of raster order and the heavy cost of multi-scale sequences.

To resolve this trade-off, we introduce \textbf{DenseAR}, an autoregressive paradigm based on \emph{next-dense-stride prediction} that combines the compact single grid of raster order with the coarse-to-fine generation of next-scale order, without the former's sequential slowness or the latter's lengthy multi-scale sequence.
Our key insight is that both global layout and fine detail already reside in one grid, so the coarse-to-fine order can come from the order in which tokens are predicted rather than from the tokenizer (see Figure~\ref{fig:stride_intuition}). DenseAR first predicts a few widely spaced tokens to fix the global structure, then fills in progressively denser strides for local detail; the tokens of each stride are predicted in parallel, so generation takes only a few steps.
Keeping every modality on one short grid---rather than a multi-scale stack---is also what makes unification practical: several modalities fit in a single sequence without the blow-up that burdens multi-scale backbones, so one model can serve many tasks. This is especially valuable in medical imaging, where multi-contrast synthesis, cross-modal translation, and segmentation are usually handled by separate models and remain far less explored by AR than natural images. We validate \ourmethod\ on ImageNet generation and multi-contrast brain MRI, where a single model handles all three tasks at once.

In summary, our contributions are:
\begin{itemize}[leftmargin=*]
\item A new generative paradigm, \emph{next-dense-stride prediction}, that achieves coarse-to-fine generation on a single-scale latent grid, with neither the slow decoding of raster order nor the sequence inflation of next-scale models.
\item A unified autoregressive model for medical imaging that handles multi-contrast synthesis, cross-modal translation, and segmentation in one backbone---to our knowledge, the first AR model to span all three.
\item A \emph{task-conditional} classifier-free guidance scheme that controls how strongly each prediction follows its source modalities, tunable per task with no extra components.
\item Competitive results across both domains, with a full open-source release of the training and inference pipelines for the natural-image and medical settings.
\end{itemize}

\section{Background}
\label{sec:background}

\subsection{Visual Tokenization}
\label{sec:bg-tokenization}
Most autoregressive (AR) visual models operate not on raw pixels but on discrete tokens. A vector-quantized tokenizer such as the vector-quantized variational autoencoder (VQ-VAE)~\cite{Oordetal2017vqvae}.
For example, VQ-VAE's encoder maps an input image of size $H \times W \times 3$ to a lower-resolution feature map of size $h \times w \times d$ (with $h = H/f$, $w = W/f$ for a downsampling factor $f$), and each $d$-dimensional feature vector is replaced by its nearest entry in a learned codebook, yielding an $h \times w$ grid of token indices. This grid is what the most visual AR model predicts over.

Visual autoregressive modeling (VAR)~\cite{var, Infinity} newly introduces a multi-scale tokenizer to enable a coarse-to-fine representation. Instead of a single-scale grid per image, it encodes each image into $K$ grids $R = (\r_1, \dots, \r_K)$, where each $\r_k$ is of size $h_k \times w_k$ and the resolutions grow from coarse to fine, with the last grid $\r_K$ matching the full $h \times w$. Reconstruction sums all $K$ grids before a single decode, $\texttt{Decode}\!\left(\sum_{k=1}^{K} \texttt{Expand}_k(\r_k)\right)$,
where $\texttt{Expand}_k(\cdot)$ takes the index grid $\r_k \in \{1,\dots,V\}^{h_k \times w_k}$, looks up its codebook vectors to obtain a feature grid of size $h_k \times w_k \times d$, upsamples it to the shared resolution $h \times w \times d$, and applies a per-scale convolution; $\texttt{Decode}(\cdot)$ then maps the summed feature grid back to an image. The reconstruction therefore emerges only from the entire stack, not any single grid. Because each image is now represented by $K$ stacked grids rather than one, any model predicting over this representation must process their \emph{cumulative} length $\sum_{k=1}^{K} h_k \times w_k$---far more tokens than the single $h \times w$ grid of a single-scale tokenizer.

\begin{figure*}[t]
\begin{center}
\includegraphics[width=1.\textwidth]{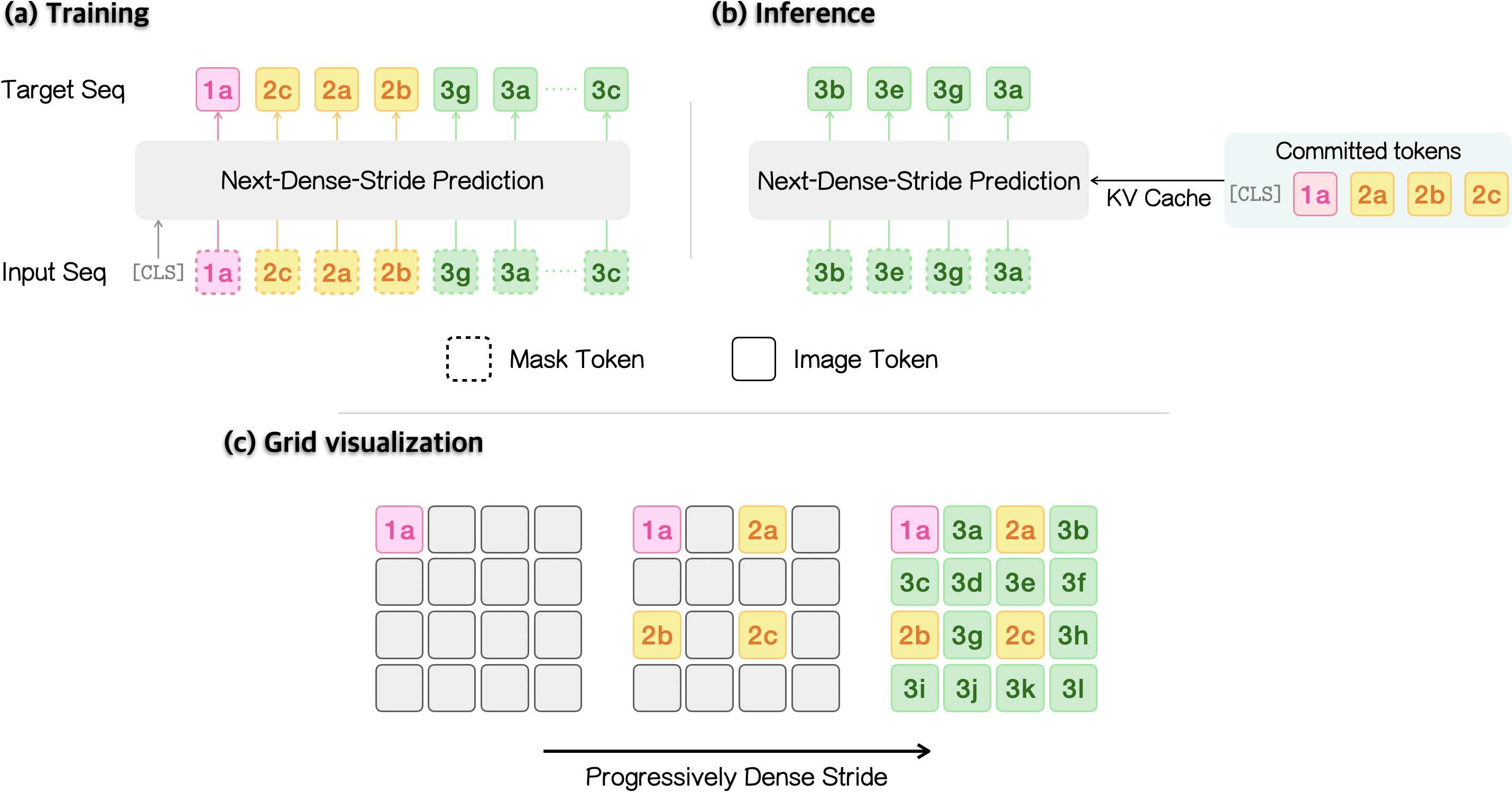}
\end{center}
\caption{Visual illustration of \ourmethod's next-dense-stride prediction. \textbf{(a) Training:} after the fixed \texttt{\textcolor{gray}{[CLS]}} prefix, each target token is predicted from a position-encoded \emph{mask-token query} (dashed) at that position.
Queries are ordered by stride---stage~1 (\textcolor{pink}{pink}), stage~2 (\textcolor{YellowOrange}{yellow}), stage~3 (\textcolor{Green}{green}).
Stages run coarse-to-fine, but tokens \emph{within} a stage are randomly ordered (e.g.\ \textcolor{YellowOrange}{\texttt{2c, 2a, 2b}}).
\textbf{(b) Inference:} from \texttt{\textcolor{gray}{[CLS]}}, each step predicts the next group of tokens in parallel using their mask-token queries, conditioned on the committed context. That context is held in a key-value (KV) cache, so predicted tokens are encoded once and reused without recomputation.
\textbf{(c) Grid visualization:} the stride densifies left to right---coarse positions fill first, then denser stages complete the grid.
Note that the three stages shown here are illustrative, based on a $4\times4$ grid example. The number of stages varies with the tokenizer's grid size.
}
\vspace{-.3cm}
\label{fig:token_illustration}
\end{figure*}

\subsection{Autoregressive Generation and Token Ordering}
\label{sec:bg-ordering}
Given a token grid, an AR model factorizes the joint distribution over its tokens and predicts them one at a time,
\begin{equation}
    p(\x_1, \x_2, \dots, \x_T) = \prod_{t=1}^{T} p(\x_t \mid \x_1, \x_2, \dots, \x_{t-1}).
\end{equation}
The order in which tokens are visited is a design choice, and it is where visual AR methods differ most.

\paragraph{Single-scale order.}
Single-scale methods predict over one $h \times w$ grid. The standard choice is raster order, as in LlamaGen~\cite{sun2024Llamagen}, which flattens the grid into $hw$ tokens and predicts them one at a time with causal self-attention---hundreds of sequential steps per image, hence slow inference. A line of work keeps the single grid but predicts tokens in parallel to cut this step count: the masked generative image transformer (MaskGIT)~\cite{maskgit} decodes subsets of positions simultaneously via masked prediction.
The parallelized autoregressive model (PAR)~\cite{wang2024par} groups spatially distant, weakly dependent tokens for simultaneous prediction to accelerate inference.
The autoregressive model with randomized parallel generation (ARPG)~\cite{li2026arpg} trains causally on randomized token orders and, at inference, uses position-encoded mask query tokens to predict tokens in any order in parallel, caching each committed group so it is encoded once and reused as context rather than recomputed at every step.

\paragraph{Multi-scale order.}
To combine parallel decoding with a coarse-to-fine order, VAR~\cite{var} changes the prediction unit from a single token to one entire token map per scale. Using the $K$ grids from Section~\ref{sec:bg-tokenization}, it factorizes generation across scales,
\begin{equation}
    p(\r_1, \r_2, \dots, \r_K) = \prod_{k=1}^{K} p(\r_k \mid \r_1, \r_2, \dots, \r_{k-1}),
\end{equation}
predicting all tokens within a grid $\r_k$ in parallel and cutting the step count to the number of scales $K$. This coarse-to-fine prior is not free: as noted in Section~\ref{sec:bg-tokenization}, the model processes over the full stack of grids rather than the final $h \times w$ grid alone, so the larger token budget translates directly into higher memory and attention compute, making these models expensive to scale, especially in multitask settings.

The two families trade off differently: single-scale orders keep one compact grid---and can even decode it in parallel---but offer no coarse-to-fine prior, while multi-scale orders are coarse-to-fine and parallel yet inflate the token budget. DenseAR introduces a third option---a coarse-to-fine order obtained \emph{on the single grid} by varying the prediction stride, retaining parallel decoding without the multi-scale inflation.

\subsection{Unified Generative Models for Multimodal and Multi-task Vision}
\label{sec:unified-related-work}
Beyond single-task generation, a growing body of work builds \emph{unified} models spanning multiple modalities and tasks within one generative model. We review how diffusion- and autoregressive-based approaches address multimodal and multi-task problems across natural and medical images.

\begin{figure*}[t]
\begin{center}
\includegraphics[width=1.\textwidth]{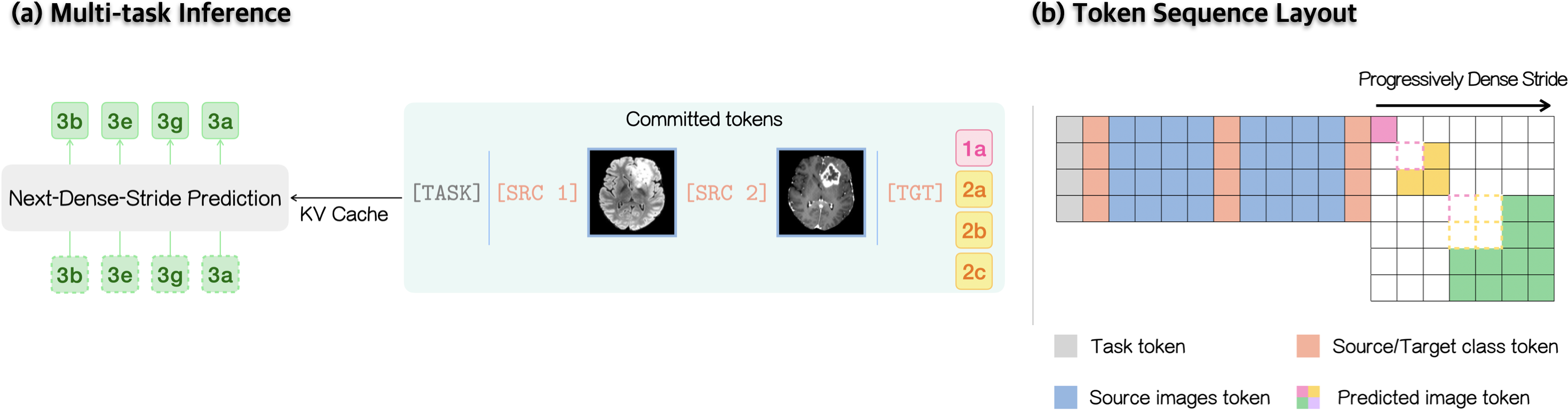}
\end{center}
\caption{\textbf{Unified multi-task inference.} \textbf{(a) Multi-task inference.} A task is one token sequence: a task marker (\textcolor{gray}{\texttt{[TASK]}}), zero or more source modalities each with a class marker (\textcolor{Melon}{\texttt{[SRC]}}), and a target after \textcolor{Melon}{\texttt{[TGT]}}. This prefix plus the target tokens decoded so far forms the committed context, which the model reads from the key--value (KV) cache to predict the next group of target tokens in parallel; token colors match \textbf{(b)}. \textbf{(b) Token sequence layout.} Each square is one token, ordered left to right. The prefix---task token (\textcolor{gray}{gray}), class tokens (\textcolor{Melon}{orange}), and source-image tokens (\textcolor{blue}{blue})---is a fixed context; only the target tokens (right) are generated, coarse to fine. The model first commits the sparse token (\textcolor{pink}{pink}), then progressively denser groups (\textcolor{YellowOrange}{yellow}, \textcolor{Green}{green}), so the colored target region grows toward the right.
Source modalities are drawn as images for readability; in the sequence, they are discrete tokens.}
\vspace{-.3cm}
\label{fig:multimodal_token_illustration}
\end{figure*}

\paragraph{Diffusion-based unification.}
Diffusion-based unification generally follows two paradigms. The first incorporates separate adapters or conditioning modules for each distinct condition type~\cite{zhang2023controlnet, mou2023t2i, huang2023composer}. The second relies on a single shared backbone across tasks~\cite{liu2025univg}, though often requiring task- or modality-specific encoders to apply each condition. Similar architectural strategies have emerged in medical imaging.
For instance, medical multi-modal generation (MedM2G)~\cite{zhan2024medm2g}, which trains a separate diffusion model per modality and couples them via a central text latent space and coupled through shareable cross-attention mechanisms.
Similarly, Adaptive LDM~\cite{kim2024adaptive} integrates switchable, target-specific normalization blocks to translate a single source modality into various targets.
Both approaches integrate modality-specific modules directly into the network architecture.
In contrast, Any2all~\cite{gan2026unified} instead stacks modalities as input channels, avoiding per-modality components but fixing the set of modalities in advance, so adding one requires retraining the whole model.

\paragraph{Autoregressive-based unification.}
Autoregressive models extend to new tasks by adding conditioning to the token sequence rather than new per-task modules.
EditAR~\cite{mu2025editAR} casts editing and translation as next-token prediction, but depends on an external text encoder and a foundation-model distillation module to align its predictions.
VAREdit~\cite{mao2025varedit} instead casts editing as next-scale prediction on a pretrained multi-scale VAR backbone.
The full multi-scale source is too costly to use, so VAREdit keeps only the finest scale and adds a module to compensate. Yet the target mainly remains a cumulative multi-scale stack, so the cost persists.
Medical efforts remain single-task: MedVAR~\cite{he2026medvar} adapts VAR to medical synthesis, and AR-Seg~\cite{chen2025arseg} applies next-scale prediction to segmentation, both inheriting VAR's multi-scale sequence.
To our knowledge, no prior AR backbone handles multi-contrast synthesis, cross-modal translation, and segmentation in one inference-efficient model.
\ourmethod\ fills this gap: one backbone, no text encoder, both continuous images and discrete segmentation maps, with coarse-to-fine generation from stride scheduling on a single grid.

Outside of diffusion and autoregressive pipelines, medical vision generalist (MVG)~\cite{ren2024MVG} unifies multiple image-to-image generation tasks (such as cross-modal translation, segmentation, and inpainting) using in-context learning.
To be specific about in-context learning, MVG specifies each task through an example pair: a source image and its ground-truth target, such as a scan and its segmentation mask, or a source-modality scan and the matching target-modality scan. This pair is fed alongside the test image, the input for which we actually want a prediction. MVG stitches the example pair and the test image into one canvas, then fills in the held-out target with a masked vision transformer~\cite{dosovitskiy2020vit} under a regression loss. Two drawbacks follow. First, the example pair is the only task specification, so every prediction needs a fully labeled pair at inference.
Second, the single canvas holds one source image, so MVG cannot condition on several test inputs at once.

\section{Method}
\label{sec:method}
\ourmethod\ performs coarse-to-fine autoregression on a \emph{single} latent grid through \emph{next-dense-stride prediction}. It generates an image stride by stride: a few widely spaced tokens are committed first to fix the global layout, then progressively denser strides fill in detail. At inference the stride order lets the model predict many tokens in parallel at each step, so a full image takes only a few steps rather than the token-by-token raster-order generation. This gives a coarse-to-fine prior like that of next-scale models, but on one compact grid and without their long multi-scale token sequences. We define the stride order (Section~\ref{sec:method-order}) and then extend it into a single backbone for unified multimodal, multi-task modeling (Section~\ref{sec:method-unified}).

\subsection{Next-dense-stride Order}
\label{sec:method-order}
A tokenizer maps an image to a grid $\x\in\{1,\dots,V\}^{G\times G}$ of $G^2$ discrete tokens. \ourmethod\ visits these tokens from coarse to fine along a \emph{stride pyramid}. At stride $k$, we keep every $k$-th token along each axis, giving a sub-grid $D_k$ of $(G/k)^2$ evenly spaced tokens---for example, on a $16\times16$ grid, stride $k=8$ keeps a $2\times2$ sub-grid of $4$ tokens, while stride $k=1$ keeps all $256$. The strides form a decreasing sequence $K,\,K/2,\,\dots,\,1$, where $K$ is the coarsest stride. Halving the stride nests the sub-grids,
\begin{equation}
D_{K}\subset D_{K/2}\subset\cdots\subset D_1,
\end{equation}
with $D_1$ the full grid. Generation walks down this sequence: it commits all of $D_{K}$, then repeatedly halves the stride, each time adding only the positions not already covered by a coarser stride, until the grid is complete (Figure~\ref{fig:token_illustration}). Choosing $K=G$ starts from a single token; a smaller $K$ starts from a denser sub-grid.

Each halving thus defines a \emph{stage}---the set of positions newly added at that stride---and the stages run strictly coarse to fine.
The tokens \emph{within} a stage, however, have no inherent order, so we visit them in a random within-stage order (Figure~\ref{fig:token_illustration}).

\paragraph{Training.}
\ourmethod\ uses the same maximum-likelihood objective as a standard autoregressive model, only over a sampled stride order $\pi$ rather than a fixed raster one.
For each image, we draw an order $\pi$: we pick the coarsest stride $k_0$ from a fixed set, then shuffle the tokens within each stage.
Following the order $\pi$, the model predicts each token conditioned on those preceding it,
\begin{equation}
\mathcal{L} = -\,\mathbb{E}_{\x,\,\pi}\!\left[\sum_{i=1}^{G^2}\log p_\theta\!\big(\x_{\pi_i}\mid \x_{\pi_{<i}}\big)\right].
\end{equation}
We shuffle within each stage so the model learns to predict any position in the next stage given any coarser context. Training, therefore, covers a distribution of stride orders rather than a single fixed path, and this is exactly what lets inference decode an entire stage in parallel: whatever grouping we choose at test time, the model has been trained to predict it from the committed coarser tokens.
Despite the multi-stage coarse-to-fine design, every stage stays on the same $G\times G$ grid, so the conditioning sequence never exceeds $G^2$---unlike next-scale order.

\paragraph{Inference.}
At inference, \ourmethod\ follows the same coarse-to-fine order, but decodes many tokens per step rather than one at a time.
Denoting $\x_{<t}$ for the tokens committed in earlier steps and $P_t$ for the group predicted at step $t$, the group is predicted in parallel,
\begin{equation}
p_\theta\big(\x_{P_t}\mid \x_{<t}\big) \;=\; \prod_{j\in P_t} p_\theta\big(\x_j \mid \x_{<t}\big),
\end{equation}
where every token in the step is drawn from the same committed context $\x_{<t}$, not from the other tokens in $P_t$.
Training is fully token-causal, with each token conditioning on all earlier tokens in the order; inference relaxes only this, predicting a whole group $P_t$ in parallel so its tokens do not condition on one another.
The random within-stage order seen during training is what makes this parallel grouping safe.
Committed tokens are stored in a key-value (KV) cache and reused across steps, so context is encoded once rather than recomputed as in Figure~\ref{fig:token_illustration}, and the number of tokens in $P_t$ follows a schedule chosen at inference (detail is in Section~\ref{sec:numericalevaluations}).

\subsection{Unified Multimodal, Multi-task Model}
\label{sec:method-unified}
We extend \ourmethod\ to a unified model that handles multiple modalities and tasks within one backbone.
Prior unified models add structure for each task: diffusion approaches attach a new encoder or conditioning module per task~\cite{kim2024adaptive, zhan2024medm2g}, and autoregressive approaches inherit the long multi-scale sequence of next-scale tokenizers and stack further conditioning and refinement modules on top of it~\cite{qu2025VARSR, rajagopalan2025restorevar}.
\ourmethod\ needs none of these.
Its coarse-to-fine order comes from the prediction stride on a single grid, so it avoids the multi-scale sequence growth that is especially costly when several modalities share one sequence.
Because every modality shares the same token domain, a task is simply a choice of which grids are given and which are predicted, expressed entirely in the token sequence with a small modality marker prepended to each grid.
Concretely, each modality is a grid of discrete vector-quantized (VQ) tokens $\x^m\in\{1,\dots,V\}^{G\times G}$, where $m$ denotes the modality (e.g.\ $m=\texttt{seg\_mask}$ for a segmentation mask). A task predicts one target grid from a set of source grids $S$, with each grid preceded by a modality marker.
A task is written as one token sequence with three parts,
\begin{equation}
\begin{aligned}
\z = \big[\,
&\underbrace{\texttt{[TASK]}}_{\text{task marker}}\;\\
&\underbrace{\texttt{[SRC 1]}\;\x^{1}\;\texttt{[SRC 2]}\;\x^{2}\dots}_{\text{sources (raster)}}\;\\
&\underbrace{\texttt{[TGT]}\;\x^{\star}}_{\text{target (stride order }\pi)}\,\big].
\end{aligned}
\end{equation}
where each modality grid is preceded by a marker: a task marker, a \texttt{[SRC]} marker per source modality (laid out in raster order, since sources are fixed context rather than generated), and a \texttt{[TGT]} marker for the target grid $\x^{\star}$, the only portion generated, in stride order $\pi$.
Tasks differ only in the task marker \texttt{[TASK]} and the source set $S$, drawn from the same sequence (Figure~\ref{fig:multimodal_token_illustration}).
With $\texttt{[TASK] = generate}$, $S = \varnothing$, and $\texttt{[TGT] = FLAIR}$, the model generates a \texttt{FLAIR} image; with $\texttt{[TASK] = translate}$ and $S$ holding the available source contrasts, it predicts the same \texttt{FLAIR} target conditioned on those sources.

\paragraph{Task-conditional classifier-free guidance.}
The same design supports classifier-free guidance (CFG)~\cite{cfg} with no extra components.
Standard CFG is widely used in class-conditional image generation, where it trades off sample fidelity against diversity by guiding generation toward a class label.
We extend it to a task-conditional form that instead controls how strongly the target follows its source modalities, separately for each task.
This is useful because the right amount of source adherence differs across tasks: translation should stay close to the source contrast, whereas modality-conditioned generation should follow the task prior more loosely.
During training, we replace the source blocks with dummy blocks of the same length with probability $p_{\text{drop}}$, while keeping the task marker, so the model also learns a source-free prior $p_\theta(\x^{\star}\mid\texttt{[TASK]})$ that is still specific to the task.
At inference, we combine the source-conditioned and source-free predictions for each token, with guidance strength $w$,
\begin{equation}
\begin{split}
\log \tilde{p}_\theta \;=\;& (1+w)\,\log p_\theta\big(\cdot \mid \texttt{[TASK]}, S\big) \\
&-\; w\,\log p_\theta\big(\cdot \mid \texttt{[TASK]}\big),
\end{split}
\end{equation}
so guidance steers toward the sources while staying within the task. We train all tasks jointly under the objective above, drawing each task from its own data and mixing tasks across steps; the task mixture, guidance strength $w$, source-dropout rate, and data splits are given in Section~\ref{sec:numericalevaluations}.

\begin{table}[b]
\centering
\caption{Capability versus required components among generative models. The top block marks the tasks each model addresses; the bottom block, the components it requires. \ourmethod\ covers all three tasks with the leanest requirements: a single VQ-VAE, no text encoder, no pretrained backbone, and no in-context example. \emph{In-context example} denotes a labeled pair (input plus ground-truth output) supplied at inference to specify the task. \cmark\,= supported/required, \xmark\,= not.}
\label{tab:requirements}
\renewcommand{\arraystretch}{1.3}
\setlength{\tabcolsep}{8pt}
\resizebox{\linewidth}{!}{%
\begin{tabular}{l|c|c|c|c|c}
\hline 
\multicolumn{1}{l}{} & \multicolumn{1}{c}{LlamaGen} & \multicolumn{1}{c}{DiT} & \multicolumn{1}{c}{EditAR} & \multicolumn{1}{c}{MVG} & \multicolumn{1}{c}{\textbf{Ours}} \\
\multicolumn{1}{l}{} & \multicolumn{1}{c}{\cite{sun2024Llamagen}} & \multicolumn{1}{c}{\cite{peebles2023dit}} & \multicolumn{1}{c}{\cite{mu2025editAR}} & \multicolumn{1}{c}{\cite{ren2024MVG}} & \\
\hline 
\multicolumn{6}{c}{\cellcolor{gray!12}\textit{\textbf{Capability}}} \\
Generation        & \cmark & \cmark & \cmark & \xmark & \cmark \\
Translation       & \xmark & \xmark & \cmark & \cmark & \cmark \\
Segmentation      & \xmark & \xmark & \xmark & \cmark & \cmark \\
\hline 
\multicolumn{6}{c}{\cellcolor{gray!12}\textit{\textbf{Required components}}} \\
Autoencoder       & \cmark & \cmark & \cmark & \xmark & \cmark \\
Text encoder      & \xmark & \xmark & \cmark & \xmark & \xmark \\
Pretrained backbone  & \xmark & \xmark & \cmark & \cmark & \xmark \\
In-context example  & \xmark & \xmark & \xmark & \cmark & \xmark \\
\hline 
\end{tabular}%
}
\end{table}

\newcommand{\gr}[1]{\textcolor{gray}{#1}}  

\begin{table*}[t]
\centering
\caption{Unified evaluation across translation, generation, and segmentation on BraTS-2023. The left block (Capability) shows which tasks each method is designed for; the Efficiency block reports inference cost; the right block reports quantitative metrics.
A single \ourmethod\ leads on translation, ranks second on generation, and is competitive on segmentation, while being the only model that addresses all three tasks.
Efficiency is computed with a batch size of 64 and bfloat16 precision.
\hlgreen{Best} and \hlblue{second-best} results are color-coded per task among the generalist models.}
\label{tab:unified_v2}
\renewcommand{\arraystretch}{1.0}
\setlength{\tabcolsep}{5pt}
\resizebox{\textwidth}{!}{%
\begin{tabular}{l ccc c c c cccc ccc cc}
\toprule
& \multicolumn{3}{c}{\textbf{Capability}} & \phantom{a}
& \multicolumn{1}{c}{\textbf{Efficiency}} & \phantom{a}
& \multicolumn{4}{c}{\textbf{Translation}}
& \multicolumn{3}{c}{\textbf{Generation}}
& \multicolumn{2}{c}{\textbf{Segmentation}} \\
\cmidrule(lr){2-4} \cmidrule(lr){6-6} \cmidrule(lr){8-11} \cmidrule(lr){12-14} \cmidrule(lr){15-16}
Method & Trans. & Gen. & Seg. & & Mem\,(GB) & & PSNR\,$\uparrow$ & SSIM\,$\uparrow$ & LPIPS\,$\downarrow$ & L1\,$\downarrow$ & FID\,$\downarrow$ & RadFID\,$\downarrow$ & KID\,$\downarrow$ & Dice\,$\uparrow$ & IoU\,$\uparrow$ \\
\midrule
\multicolumn{16}{c}{\cellcolor{gray!7}\textbf{Specialists}} \\
\midrule
\gr{Pix2Pix}~\cite{pix2pix2017}         & \multirow{3}{*}{\cmark} & \multirow{3}{*}{\xmark} & \multirow{3}{*}{\xmark} & & \gr{0.56}  & & \gr{21.74} & \gr{0.831} & \gr{0.100} & \gr{0.035} & \gr{--} & \gr{--} & \gr{--} & \gr{--} & \gr{--} \\
\gr{CycleGAN}~\cite{CycleGAN2017}       &        &        &        & & \gr{0.92}  & & \gr{20.29} & \gr{0.819} & \gr{0.105} & \gr{0.048} & \gr{--} & \gr{--} & \gr{--} & \gr{--} & \gr{--} \\
\gr{MSG-LDM}~\cite{lin2026msgldm}       &        &        &        & & \gr{8.16}  & & \gr{21.10} & \gr{0.814} & \gr{0.112} & \gr{0.039} & \gr{--} & \gr{--} & \gr{--} & \gr{--} & \gr{--} \\
\midrule
\gr{LlamaGen}~\cite{sun2024Llamagen}    & \multirow{2}{*}{\xmark} & \multirow{2}{*}{\cmark} & \multirow{2}{*}{\xmark} & & \gr{6.43}  & & \gr{--} & \gr{--} & \gr{--} & \gr{--} & \gr{10.92} & \gr{0.011} & \gr{0.005} & \gr{--} & \gr{--} \\
\gr{DiT}~\cite{peebles2023dit}          &        &        &        & & \gr{9.11}  & & \gr{--} & \gr{--} & \gr{--} & \gr{--} & \gr{8.03} & \gr{0.005} & \gr{0.004} & \gr{--} & \gr{--} \\
\midrule
\gr{nnU-Net}~\cite{Isenseeetal2021nnUNet}    & \multirow{2}{*}{\xmark} & \multirow{2}{*}{\xmark} & \multirow{2}{*}{\cmark} & & \gr{1.30}  & & \gr{--} & \gr{--} & \gr{--} & \gr{--} & \gr{--} & \gr{--} & \gr{--} & \gr{0.898} & \gr{0.828} \\
\gr{MedSegDiff-V2}~\cite{wu2023medsegdiffv2} &        &        &        & & \gr{15.07} & & \gr{--} & \gr{--} & \gr{--} & \gr{--} & \gr{--} & \gr{--} & \gr{--} & \gr{0.851} & \gr{0.765} \\
\midrule
\multicolumn{16}{c}{\cellcolor{gray!7}\textbf{Generalists}} \\
\midrule
EditAR~\cite{mu2025editAR}   & \cmark & \cmark & \xmark & & 24.61 & & \hlblue{20.78} & 0.811 & \hlblue{0.112} & 0.041 & \hlblue{15.88} & \hlblue{0.029} & \hlblue{0.010} & -- & -- \\
\midrule
MVG~\cite{ren2024MVG}        & \cmark & \xmark & \cmark & & \hlblue{11.00}  & & 20.46 & \hlblue{0.825} & 0.114 & \hlblue{0.040} & -- & -- & -- & \hlgreen{0.872} & \hlgreen{0.790} \\
\midrule
\textbf{Ours}   & \cmark & \cmark & \cmark & & \hlgreen{10.72} & & \hlgreen{21.83} & \hlgreen{0.833} & \hlgreen{0.099} & \hlgreen{0.033} & \hlgreen{8.50} & \hlgreen{0.012} & \hlgreen{0.004} & \hlblue{0.851} & \hlblue{0.756} \\
\bottomrule
\end{tabular}%
}
\end{table*}

\begin{figure*}[htbp]
\begin{center}
\includegraphics[width=0.75\textwidth]{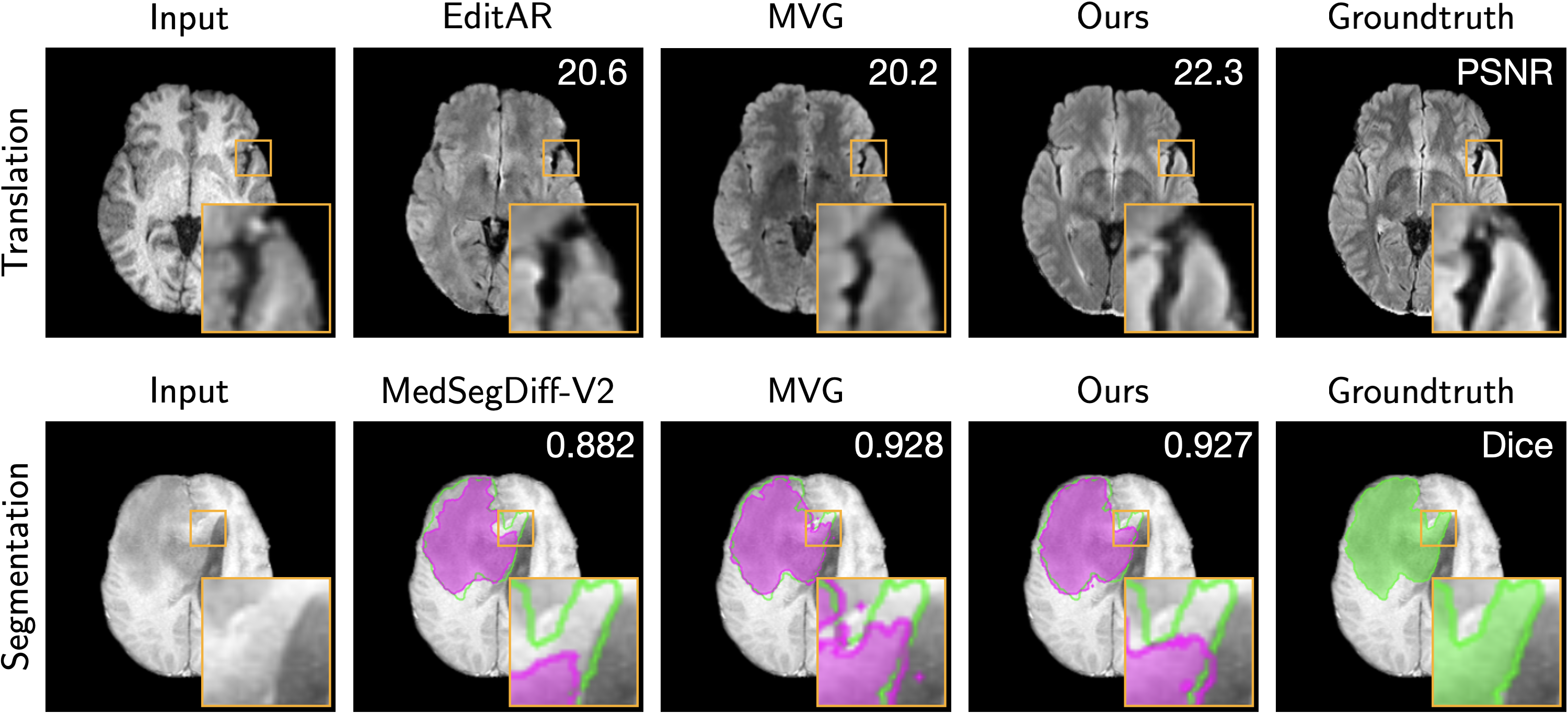}
\end{center}
\caption{Visual comparison of translation and segmentation results for representative methods from Table~\ref{tab:unified_v2}.
In segmentation, the \textcolor{maskpink}{pink mask} denotes the predicted segmentation and the \textcolor{maskgreen}{green mask} denotes the ground truth.}
\vspace{-.3cm}
\label{fig:medical_tasks}
\end{figure*}

\section{Numerical Evaluations}
\label{sec:numericalevaluations}

We evaluate \ourmethod's two contributions separately: (1) the next-dense-stride order, which delivers coarse-to-fine generation on a single grid, and (2) the unified backbone, which handles many modalities and tasks within one model. To test (1), we evaluate class-conditional image generation on ImageNet~\cite{deng2009imagenet}, the standard benchmark for visual autoregressive models across raster, masked, and next-scale orders~\cite{llamagen, Esseretal2021VQGAN, li2026arpg, wang2024par, maskgit}. To test (2), we turn to multi-contrast brain MRI, using the BraTS-2023~\cite{Menzeetal2015brats1, Bakasetal2018brats2}. Unified medical imaging today is almost entirely diffusion-based and splits one task per model; to our knowledge, \ourmethod\ is the first autoregressive model to handle multi-contrast generation, cross-modal translation, and segmentation within a single backbone. We describe datasets and training details in Section~\ref{sec:eval-setup}, baselines in Section~\ref{sec:eval-baselines}, and results in the sections that follow.

\begin{table*}[t]
\centering
\caption{Quantitative comparison of multimodal translation on BraTS 2023.
$\uparrow$/$\downarrow$: higher/lower is better. \hlgreen{Best} and \hlblue{second-best} results are color-coded per translation pair.}
\label{tab:translation_main}
\renewcommand{\arraystretch}{1.0}
\setlength{\tabcolsep}{4pt}
\setlength{\dashlinedash}{1.2pt}
\setlength{\dashlinegap}{1.5pt}
\resizebox{0.9 \textwidth}{!}{%
\begin{tabular}{ll cccc cccc cccc}
\toprule
\multirow{2}{*}{Source} & \multirow{2}{*}{Target}
& \multicolumn{4}{c}{EditAR~\cite{mu2025editAR}}
& \multicolumn{4}{c}{MVG~\cite{ren2024MVG}}
& \multicolumn{4}{c}{\textbf{Ours}} \\
\cmidrule(lr){3-6} \cmidrule(lr){7-10} \cmidrule(lr){11-14}
& & PSNR$\uparrow$ & SSIM$\uparrow$ & LPIPS$\downarrow$ & L1$\downarrow$
& PSNR$\uparrow$ & SSIM$\uparrow$ & LPIPS$\downarrow$ & L1$\downarrow$
& PSNR$\uparrow$ & SSIM$\uparrow$ & LPIPS$\downarrow$ & L1$\downarrow$ \\
\midrule
\multirow{3}{*}{T1} & T1ce & \hlblue{21.14} & \hlblue{0.825} & \hlblue{0.115} & 0.038 & 20.59 & 0.816 & 0.123 & \hlblue{0.036} & \hlgreen{22.42} & \hlgreen{0.856} & \hlgreen{0.097} & \hlgreen{0.032} \\
& T2 & 21.02 & 0.827 & 0.110 & 0.038 & \hlblue{22.27} & \hlblue{0.862} & \hlblue{0.100} & \hlblue{0.034} & \hlgreen{22.54} & \hlgreen{0.863} & \hlgreen{0.090} & \hlgreen{0.032} \\
& FLAIR & \hlblue{20.53} & 0.806 & 0.117 & 0.042 & 20.46 & \hlblue{0.825} & \hlblue{0.114} & \hlblue{0.040} & \hlgreen{21.91} & \hlgreen{0.839} & \hlgreen{0.096} & \hlgreen{0.034} \\
\hdashline
\multicolumn{2}{l}{\textbf{Average}} & 20.90 & 0.819 & 0.114 & 0.039 & \hlblue{21.11} & \hlblue{0.834} & \hlblue{0.112} & \hlblue{0.037} & \hlgreen{22.29} & \hlgreen{0.853} & \hlgreen{0.094} & \hlgreen{0.033} \\
\midrule
\multirow{3}{*}{T1ce} & T1 & 21.38 & \hlblue{0.837} & \hlblue{0.106} & \hlblue{0.036} & \hlblue{21.44} & 0.835 & 0.118 & 0.044 & \hlgreen{23.19} & \hlgreen{0.874} & \hlgreen{0.086} & \hlgreen{0.029} \\
& T2 & 20.80 & 0.818 & 0.112 & 0.038 & \hlblue{22.36} & \hlgreen{0.862} & \hlblue{0.098} & \hlblue{0.034} & \hlgreen{22.38} & \hlblue{0.855} & \hlgreen{0.092} & \hlgreen{0.032} \\
& FLAIR & \hlblue{20.30} & 0.800 & \hlblue{0.120} & \hlblue{0.043} & 19.75 & \hlblue{0.805} & 0.125 & 0.044 & \hlgreen{21.73} & \hlgreen{0.834} & \hlgreen{0.100} & \hlgreen{0.035} \\
\hdashline
\multicolumn{2}{l}{\textbf{Average}} & 20.83 & 0.818 & \hlblue{0.113} & \hlblue{0.039} & \hlblue{21.18} & \hlblue{0.834} & 0.114 & 0.041 & \hlgreen{22.43} & \hlgreen{0.854} & \hlgreen{0.093} & \hlgreen{0.032} \\
\midrule
\multirow{3}{*}{T2} & T1 & 20.69 & 0.833 & 0.115 & \hlblue{0.039} & \hlblue{21.59} & \hlblue{0.861} & \hlblue{0.110} & 0.044 & \hlgreen{22.50} & \hlgreen{0.871} & \hlgreen{0.092} & \hlgreen{0.031} \\
& T1ce & \hlblue{20.50} & 0.812 & 0.125 & \hlblue{0.042} & 19.99 & \hlblue{0.835} & \hlblue{0.124} & 0.051 & \hlgreen{21.93} & \hlgreen{0.846} & \hlgreen{0.105} & \hlgreen{0.034} \\
& FLAIR & 20.41 & 0.800 & 0.120 & 0.043 & \hlblue{21.03} & \hlgreen{0.843} & \hlblue{0.109} & \hlblue{0.038} & \hlgreen{22.20} & \hlblue{0.842} & \hlgreen{0.097} & \hlgreen{0.033} \\
\hdashline
\multicolumn{2}{l}{\textbf{Average}} & 20.53 & 0.815 & 0.120 & \hlblue{0.041} & \hlblue{20.87} & \hlblue{0.846} & \hlblue{0.114} & 0.044 & \hlgreen{22.21} & \hlgreen{0.853} & \hlgreen{0.098} & \hlgreen{0.033} \\
\midrule
\multirow{3}{*}{FLAIR} & T1 & 20.36 & 0.813 & \hlblue{0.116} & \hlblue{0.041} & \hlblue{21.60} & \hlblue{0.844} & 0.116 & 0.045 & \hlgreen{22.32} & \hlgreen{0.857} & \hlgreen{0.092} & \hlgreen{0.032} \\
& T1ce & \hlblue{20.28} & 0.799 & \hlblue{0.127} & \hlblue{0.043} & 19.20 & \hlblue{0.800} & 0.130 & 0.056 & \hlgreen{21.72} & \hlgreen{0.836} & \hlgreen{0.106} & \hlgreen{0.034} \\
& T2 & 20.41 & 0.804 & 0.116 & 0.041 & \hlblue{22.05} & \hlgreen{0.855} & \hlblue{0.102} & \hlblue{0.035} & \hlgreen{22.33} & \hlblue{0.851} & \hlgreen{0.092} & \hlgreen{0.032} \\
\hdashline
\multicolumn{2}{l}{\textbf{Average}} & 20.35 & 0.805 & 0.120 & \hlblue{0.042} & \hlblue{20.95} & \hlblue{0.833} & \hlblue{0.116} & 0.045 & \hlgreen{22.12} & \hlgreen{0.848} & \hlgreen{0.097} & \hlgreen{0.033} \\
\bottomrule
\end{tabular}%
}
\end{table*}

\begin{table}[b]
\centering
\caption{Effect of stride ordering on \ourmethod\ for generation on BraTS-2023. \emph{Without stride} uses random order on the same grid, with all else identical. Stride improves every metric. \hlgreen{Best} per column.}
\label{tab:stride_ablation}
\renewcommand{\arraystretch}{1.2}
\setlength{\tabcolsep}{8pt}
\begin{tabular}{l ccc}
\toprule
Method & FID\,$\downarrow$ & RadFID\,$\downarrow$ & KID\,$\downarrow$ \\
\midrule
\textbf{Ours} (without stride) & 10.54 & 0.016 & 0.006 \\
\textbf{Ours} (with stride)  & \hlgreen{8.50} & \hlgreen{0.012} & \hlgreen{0.004} \\
\bottomrule
\end{tabular}
\end{table}

\subsection{Experimental Setup}
\label{sec:eval-setup}
\paragraph{Datasets.} For medical imaging, we use BraTS-2023~\cite{Menzeetal2015brats1, Bakasetal2018brats2}, which provides four co-registered MRI contrasts per subject---native T1-weighted (T1), contrast-enhanced T1-weighted (T1ce), T2-weighted (T2), and T2 fluid-attenuated inversion recovery (FLAIR)---with an expert tumor segmentation. We take axial slices and keep the central $20$--$80\%$ of each volume, discarding near-empty end slices so every retained slice contains substantial brain anatomy. Each slice is center-cropped from $240\times240$ to $228\times228$ (verified to remove no brain tissue across all slices), resized to $256\times256$, and encoded to a $16\times16$ token grid, matching the natural-image setting. Because the official test set has no released segmentation masks, we re-split the $1{,}251$ subjects into $1{,}000$ for training ($93{,}000\times4$ slices) and $240$ for testing ($22{,}320\times4$ slices), with $11$ held out for validation ($1{,}023\times4$ slices). For the segmentation task, we exclude slices with an empty tumor mask.

For natural images, we use ImageNet~\cite{deng2009imagenet} at $256\times256$, covering 1,000 classes over 1.28 M training images; following common practice, we train class-conditional generation on an encoded image, which is in a $16\times16$ grid of VQ tokens~\cite{Oordetal2017vqvae}.

\paragraph{Implementation.} For natural images we tokenize with the pretrained VQ-VAE of LlamaGen~\cite{llamagen}.
For the medical setting, we train two VQ-VAEs of identical architecture: an image-only tokenizer that encodes the four MRI contrasts, used by all autoregressive baselines, and one that additionally encodes the tumor mask into the same shared codebook, used by \ourmethod. We thus treat the segmentation mask as just another image, so segmentation needs no segmentation-specific tokenizer or architecture.
For natural images, we train \ourmethod\ at two model sizes (L and XL); for the medical setting, we train a single unified model spanning translation, generation, and segmentation.
All model scales, mixture weights, dropout and guidance strengths, step budget, and optimization settings are listed in the appendix.

\subsection{Baselines}
\label{sec:eval-baselines}

\paragraph{Medical baselines.} On BraTS, we compare \ourmethod\ per task against strong task-specific specialists from each dominant paradigm (GAN, diffusion, autoregressive), and against the closest unified autoregressive model.
For multi-contrast generation, we use the autoregressive LlamaGen~\cite{sun2024Llamagen} and the diffusion model DiT~\cite{peebles2023dit}.
For cross-modal translation, we use the GAN-based translators Pix2Pix~\cite{pix2pix2017} and CycleGAN~\cite{CycleGAN2017}, and MSG-LDM~\cite{lin2026msgldm}, a latent-diffusion model that performs any-to-any multi-contrast translation with a single model---covering the GAN and diffusion families.
For segmentation, we use nnU-Net~\cite{Isenseeetal2021nnUNet}, the standard end-to-end medical segmentation model, and MedSegDiff-V2~\cite{wu2023medsegdiffv2}, a diffusion segmentation model.
We additionally compare to two unified models.
EditAR~\cite{mu2025editAR}, the closest unified autoregressive model, addresses translation and generation but has no native segmentation head.
MVG~\cite{ren2024MVG} is an in-context unified model addressing translation and segmentation; since it cannot condition on multiple source images, we include it only in the single-source experiments, not the multi-source translation task.
All medical baselines are trained on BraTS following each method's standard setup, with training configurations listed in Appendix~\ref{appendix:implementation_details}.
Table~\ref{tab:requirements} summarizes which tasks each model supports and the components it requires.

\begin{table*}[t]
\centering
\caption{Quantitative comparison of different methods on ImageNet benchmarks. Up ($\uparrow$) and down ($\downarrow$) arrows indicate higher and lower values are better, respectively. Efficiency was computed with a batch size of 64 and bfloat16 precision.}
\vspace{-4pt}
\label{tab:main}
\resizebox{0.62\linewidth}{!}{
\small
    \renewcommand{\arraystretch}{1.25} 
    \begin{tabular}{c|l|cr|r|cc|cc}
    \hline 
        \multirow{2}{*}{\textbf{Type}}                 & \multirow{2}{*}{\textbf{Model}} & \multirow{2}{*}{\textbf{Param.}}   & \multirow{2}{*}{\textbf{Steps}} & \textbf{Mem.} & \multirow{2}{*}{\textbf{FID\(\downarrow\)}} & \multirow{2}{*}{\textbf{IS\(\uparrow\)}}    & \multirow{2}{*}{\textbf{Pre.\(\uparrow\)}} & \multirow{2}{*}{\textbf{Rec.\(\uparrow\)}} \\
        & & & & (GB)\(\downarrow\) & & & & \\
    \hline 
        \multirow{2}{*}{Diffusion}
                                   & DiT-L/2~\cite{peebles2023dit} & 458 M & 250 & 1.62 &  5.02 & 167.2 & 0.75 & 0.57  \\
                                   & DiT-XL/2~\cite{peebles2023dit}& 675 M & 250 & 2.14 & 2.27 & 278.2 & 0.83 & 0.57  \\
        \hline 
        \multirow{3}{*}{Mask}
                                 & MaskGIT~\cite{maskgit}    & 227 M & 8   & 1.71 & 6.18 & 182.1 & 0.80    & 0.51     \\
                                 & MAR-B~\cite{mar}     & 208 M & 100 & 1.47 & 2.31 & 281.7 & 0.82 & 0.57  \\
                                 & MAR-L~\cite{mar}     & 479 M & 100 & 2.32 & 1.78 & 296.0 & 0.81 & 0.60  \\
        \hline
        \multirow{10}{*}{\shortstack{AR}}
                            & RAR-B~\cite{rar}     & 261 M & 256 & 4.65 & 1.95 & 290.5 & 0.82 & 0.58 \\
                            & RAR-L~\cite{rar}     & 461 M & 256 & 6.37 & 1.70 & 299.5 & 0.81 & 0.60 \\
                            
                            & PAR-L~\cite{wang2024par} & 343 M & 147 & 10.25 & 3.76 & 218.9 & 0.81 & 0.60 \\
                            & PAR-XL~\cite{wang2024par}& 775 M & 147 & 17.13 & 2.61 & 259.2 & 0.80 & 0.62 \\
                            \cline{2-9} 
                            & NAR-L~\cite{he2025nar} & 372 M & 31 & 10.25 & 3.06 & 263.9 & 0.81 & 0.53 \\
                            & NAR-XL~\cite{he2025nar}& 816 M & 31 & 17.13 & 2.70 & 277.5 & 0.81 & 0.58 \\
                            \cline{2-9} 
                            & RandAR-L~\cite{pang2024randar}    & 343 M & 88 &  7.32 & 2.55 & 288.8 & 0.81 & 0.58 \\
                            & RandAR-XL~\cite{pang2024randar}   & 775 M & 88 & 13.52 & 2.25 & 317.8 & 0.80 & 0.60 \\
                            \cline{2-9} 
                            & ARPG-L~\cite{li2026arpg}       & 320 M & 64  & 2.64 & 2.37 & 293.7 & 0.82 & 0.55 \\
                            
                            & ARPG-XL~\cite{li2026arpg}     & 719 M &  64 & 4.57 & 1.99 & 340.6 & 0.80 & 0.61 \\
            \hline 
        \multirow{2}{*}{Raster AR}
                              & LlamaGen-L~\cite{llamagen}   & 343 M & 576 & 10.23 & 3.07 & 256.1 & 0.83 & 0.52 \\
                              & LlamaGen-XL~\cite{llamagen}  & 775 M & 576 & 17.11 & 2.62 & 244.1 & 0.80 & 0.57 \\
        \hline
        \multirow{3}{*}{Multi-scale AR}
                              & VAR-d16~\cite{var}    & 310 M & 10 & 10.85 & 3.30 & 274.4 & 0.84 & 0.51 \\
                              & VAR-d20~\cite{var}    & 600 M & 10 & 15.97 & 2.57 & 302.6 & 0.83 & 0.56 \\
                              & VAR-d24~\cite{var}    & 1.0 B & 10 & 22.29 & 2.09 & 312.9 & 0.82 & 0.59 \\
        \hline
        \multirow{2}{*}{\ourmethod}
                            & \cellcolor{cvprblue!15}\ourmethod-L       & \cellcolor{cvprblue!15}320 M & \cellcolor{cvprblue!15}64  & \cellcolor{cvprblue!15}2.64 & \cellcolor{cvprblue!15}2.23 & \cellcolor{cvprblue!15}305.6 & \cellcolor{cvprblue!15}0.81 & \cellcolor{cvprblue!15}0.58 \\
                            
                            & \cellcolor{cvprblue!15}\ourmethod-XL       & \cellcolor{cvprblue!15}719 M &  \cellcolor{cvprblue!15}64 & \cellcolor{cvprblue!15}4.57 & \cellcolor{cvprblue!15}1.90 & \cellcolor{cvprblue!15}341.1 & \cellcolor{cvprblue!15}0.79 & \cellcolor{cvprblue!15}0.61 \\
    \hline 
    \end{tabular}
}
\vspace{-4pt}
\end{table*}

\begin{figure*}[htbp]
\begin{center}
\includegraphics[width=0.83\textwidth]{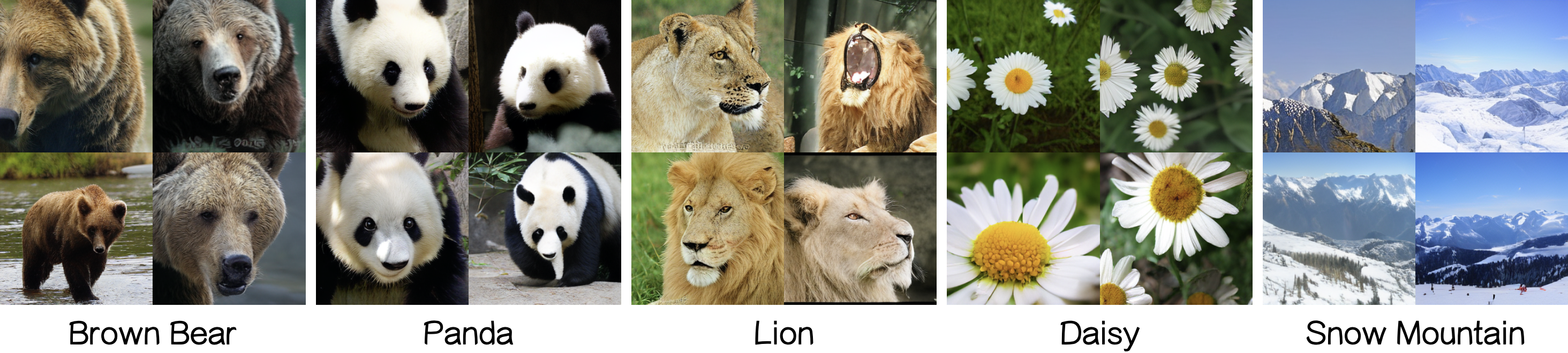}
\end{center}
\caption{Generated samples from DenseAR-XL with 64 steps. More samples are in Appendix~\ref{appendix:more_results}.}
\vspace{-.3cm}
\label{fig:class_cond_generations}
\end{figure*}

\paragraph{Natural-image baselines.} On ImageNet we compare \ourmethod\ against representative token-based visual generators covering the dominant decoding strategies---raster-order autoregression (LlamaGen~\cite{sun2024Llamagen}), masked iterative decoding (MaskGIT~\cite{maskgit}, MAR~\cite{mar}), alternative autoregressive orders (RAR~\cite{rar}, RandAR~\cite{pang2024randar}, PAR~\cite{wang2024par}, NAR~\cite{he2025nar}, ARPG~\cite{li2026arpg}), and next-scale autoregression (VAR~\cite{var})---together with the latent diffusion model DiT~\cite{peebles2023dit}.
Two comparisons are most informative. The first quantifies what the next-dense-stride paradigm contributes on its own.
By evaluating it under the same model capacity and compute as a random-order autoregressive baseline (ARPG~\cite{li2026arpg}), we attribute the resulting gains directly to coarse-to-fine stride ordering.
The second concerns coarse-to-fine modeling itself: VAR is the closest competitor, but achieves its coarse-to-fine prior by stacking token maps across resolutions, whereas \ourmethod\ achieves it as a prediction order on a single grid---so we compare not only generation quality but also the efficiency.

\paragraph{Metrics.} We report standard metrics for each task. For natural-image generation, we use Fr\'echet inception distance (FID), inception score (IS), and precision/recall, and additionally report the number of decoding steps and peak memory to quantify efficiency. For cross-modal translation, we measure fidelity to the ground-truth target with PSNR, SSIM, L1, and LPIPS. For medical generation, we report FID, the radiology-feature variant RadFID, and the kernel inception distance (KID), which better reflect perceptual quality on MRI.
For segmentation, we report the Dice coefficient and the intersection over union (IoU).

\subsection{Unified Multi-task Evaluation}
\label{sec:eval-unified}
Table~\ref{tab:unified_v2} evaluates the unified medical model on BraTS against both task-specific specialists and two unified competitors, EditAR and MVG.
The capability block tells the headline at a glance: \ourmethod\ is the only model that addresses translation, generation, and segmentation at once. Every specialist covers a single task, EditAR has no segmentation path, and MVG cannot generate from a modality label alone.
\ourmethod\ also does so with the leanest set of required components (Table~\ref{tab:requirements}): a single VQ-VAE, no text encoder, no pretrained backbone, and no in-context example---whereas EditAR relies on a text encoder and a pretrained backbone, and MVG on a pretrained backbone and a labeled in-context example.
\ourmethod\ uses the task-conditional classifier-free guidance of Section~\ref{sec:method}, with a guidance weight $w$ set separately per task; we analyze the effect of $w$ and the resulting per-task choices in Appendix~\ref{appendix:cfg-effect}.

\paragraph{Translation.} For the single direction comparable across all baselines (T1 $\to$ FLAIR, Table~\ref{tab:unified_v2}), a single \ourmethod\ backbone achieves the best cross-modal translation, improving every metric over the GAN and diffusion specialists as well as EditAR and MVG. \ourmethod\ also generalizes across all twelve source--target contrast pairs, where it generally outperforms EditAR and MVG (Table~\ref{tab:translation_main}).

\paragraph{Generation.} For each method, we generate $12{,}500$ samples per class and compute perceptual metrics over the $50{,}000$ samples against $12{,}500$ real samples per class.
In FID, DiT leads at $8.03$, \ourmethod\ follows at $8.50$, and LlamaGen is last at $10.92$.
\ourmethod\ thus ranks second on generation, with comparable KID.
Note that unlike LlamaGen and DiT, which only perform class-conditional generation, \ourmethod\ also handles translation and segmentation.

\paragraph{Segmentation.} 
For each method, we evaluate on slices whose tumor mask exceeds 
$2,000$ pixels, following the common practice of excluding near-empty slices where the metric is unstable. \ourmethod\ reaches $0.851$ Dice and $0.756$ IoU, on par with the diffusion-based MedSegDiff-V2 ($0.851$ Dice, $0.765$ IoU) and behind the dedicated specialist nnU-Net ($0.898$ Dice). \ourmethod\ recovers much of nnU-Net's accuracy through the same token interface it uses for generation and translation, with no segmentation-specific head.

Across the three tasks, one \ourmethod\ model leads on translation, ranks second on generation, and is competitive on segmentation. The breadth comes with little accuracy cost: where a specialist leads, the margin is modest, and \ourmethod\ delivers all three from a single backbone trained under one objective.

\subsection{Natural-image generation}
\label{sec:eval-natural}
Table~\ref{tab:main} reports class-conditional ImageNet generation from a \ourmethod\ model trained separately from the medical model of Section~\ref{sec:eval-unified}.
Baseline numbers for prior methods are quoted from \cite{li2026arpg}, and \ourmethod\ is run under the same evaluation protocol.
We read the results through three comparisons: raster-order autoregression (LlamaGen), which the parallel stride decoding accelerates; VAR, the closest coarse-to-fine competitor, against which we test efficiency; and a random-order baseline at matched capacity (ARPG), which isolates the contribution of the next-dense-stride order.

\paragraph{Against raster order.} Compared with raster-order autoregression, the parallel stride decoding cuts the step count sharply: \ourmethod\ generates in $64$ steps against LlamaGen's $576$, while improving FID ($2.23$ vs.\ $3.07$ at L).
More importantly, \ourmethod\ reaches this quality with both few decoding steps and low memory---a favorable quality--efficiency tradeoff that matters when many sequences are packed together in the multi-task setting of Section~\ref{sec:eval-unified}.

\paragraph{Coarse-to-fine without multi-scale inflation.} VAR obtains its coarse-to-fine prior by stacking token maps across resolutions, which inflates the sequence and the memory it occupies. \ourmethod\ reaches comparable or better quality on a single grid at a fraction of that memory: \ourmethod-XL attains $1.90$ FID using $4.57$\,GB, against VAR-d24's $2.09$ FID at $22.29$\,GB---better quality with roughly $5\times$ less memory and fewer parameters ($719$\,M vs.\ $1.0$\,B). This is the central efficiency claim of the single-grid design made concrete: the coarse-to-fine prior costs almost nothing in memory because it comes from the prediction order, not from a multi-scale tokenizer.

\paragraph{Effect of the stride order.} Because \ourmethod\ and ARPG share the same backbone, parameter count, step budget, and memory, the only difference between the two rows is the decoding order, so this pair isolates the effect of next-dense-stride prediction. At every scale \ourmethod\ improves ARPG's FID at identical cost---$2.23$ vs.\ $2.37$ at L and $1.90$ vs.\ $1.99$ at XL---while raising the inception score (e.g.\ $341.1$ vs.\ $340.6$ at XL). The stride order, therefore, adds a coarse-to-fine structure that ARPG lacks without trading away quality or efficiency.

RAR and MAR reach lower FID than \ourmethod\ ($1.70$ and $1.78$ vs.\ $1.90$ at XL). Both pay for it at decoding time. RAR emits one token per step over $256$ steps. MAR uses bidirectional attention, so each of its $100$ steps re-attends to the full sequence instead of reusing past computation. \ourmethod\ stays within $0.23$ FID of RAR-L in a quarter of the steps and at nearly $30\%$ less memory ($4.57$ vs.\ $6.37$\,GB). MAR differs further in formulation: it samples continuous tokens with a diffusion head, while \ourmethod\ keeps the plain causal next-token recipe used by unified multi-modal models.

\section{Conclusion}
\label{sec:Conclusion}

We introduce \ourmethod, an autoregressive paradigm that reformulates image generation as coarse-to-fine next-dense-stride prediction on a single latent grid. We enable the coarse-to-fine generation through the order in which tokens are predicted---a few widely spaced tokens first, then progressively denser strides---keeping the compact grid of raster order and the coarse-to-fine generation of next-scale order while avoiding the slow inference of the former and the lengthy sequences of the latter.
On class-conditional generation, \ourmethod\ is far more efficient than raster-order models at higher quality, and as strong as next-scale models using a fraction of their memory.
We further develop \ourmethod\ into a unified backbone: one \ourmethod\ model performs cross-modal translation, modality-conditioned generation, and tumor segmentation on multi-contrast brain MRI, the first medical autoregressive model to unify all three.
The key design is a single token sequence in which small markers at the front specify the task and each modality, so the model handles multiple modalities without any dedicated module to process them.
Because a task is just a choice of which grids are given and which are predicted, \ourmethod\ extends to new modalities and tasks without architectural change, such as CT-to-MRI translation or, since it operates on discrete tokens, language-guided editing. We leave these to future work.

{
\small

\bibliographystyle{IEEEbib}
\bibliography{cig}
}


\appendix

\section{Additional Results}
\label{appendix:more_results}
We provide additional results across all four tasks: modality-conditioned medical image generation, cross-modal medical image translation, medical image segmentation, and class-conditional RGB image generation.
We train two separate \ourmethod\ models, one for medical imaging and one for RGB. All medical results come from the single medical backbone, with the task selected only by the source set and modality markers; the RGB results come from the separate RGB model.

\begin{figure*}[t]
\begin{center}
\includegraphics[width=.85\textwidth]{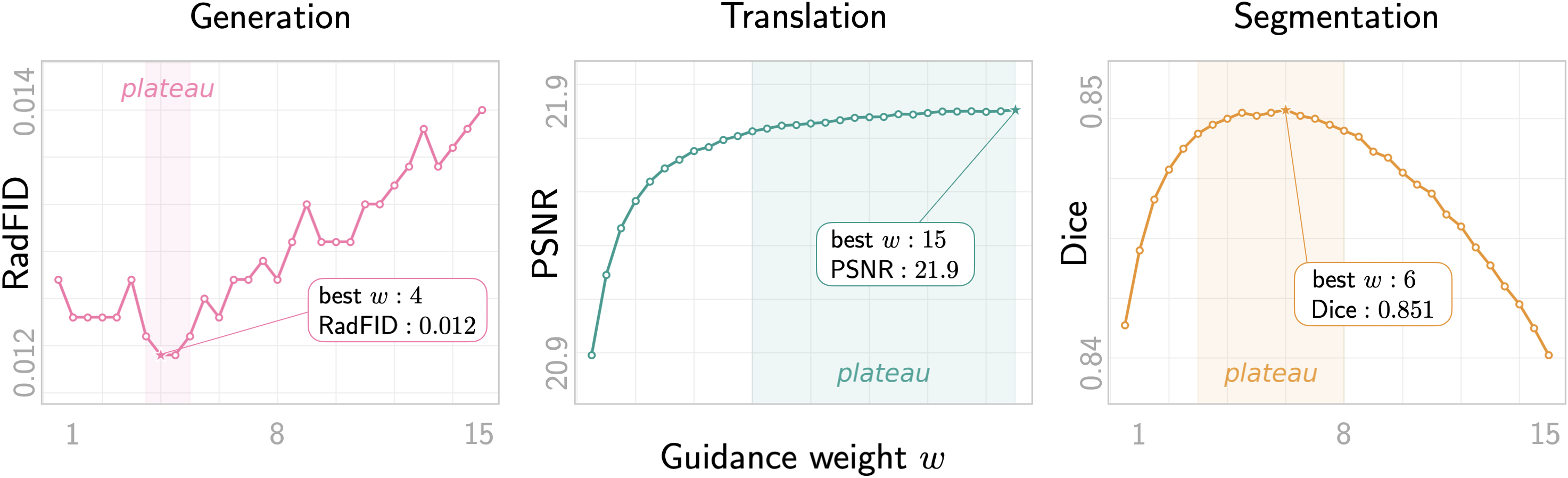}
\end{center}
\caption{Effect of the task-conditional guidance weight $w$ on each task, swept from $0.5$ to $15$. Generation is measured by RadFID ($\downarrow$), translation by PSNR ($\uparrow$), and segmentation by Dice ($\uparrow$); the best $w$ per task is starred, and shaded bands mark the plateau where the metric stays near its best. The optimal $w$ differs across tasks---low for generation ($w{=}4$), high for translation ($w{=}15$), and intermediate for segmentation ($w{=}6$)---which is why a single shared guidance strength would be suboptimal and \ourmethod\ sets $w$ per task.}

\vspace{-.3cm}
\label{fig:ablation_guidance}
\end{figure*}

\subsection{Medical Image Experiments}

\begin{figure*}[htbp]
\begin{center}
\includegraphics[width=1.0\textwidth]{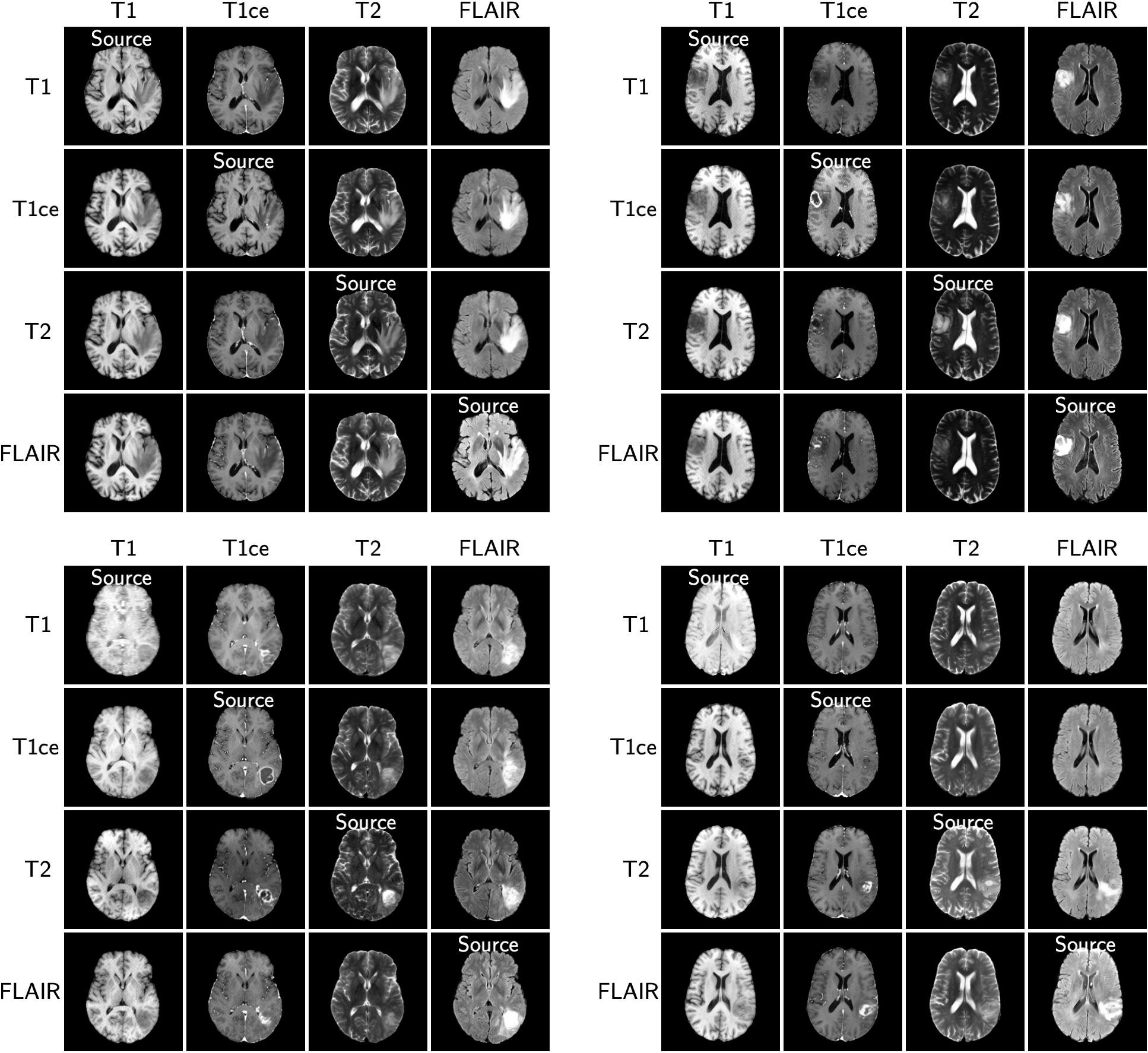}
\end{center}
\caption{\textbf{All-pairs single-source translation with a single \ourmethod\ model on BraTS-2023.} Each $4\times4$ block shows one subject: rows index the source contrast and columns the target contrast (T1, T1ce, T2, FLAIR). The cell marked \emph{Source} on the diagonal is the input given to the model; every off-diagonal cell is \ourmethod's prediction of that target contrast. A single model produces all twelve source--target translations.}
\label{fig:ablation_more_translations1}
\vspace{-.3cm}
\end{figure*}

\paragraph{Cross-modal MRI translation.} Figure~\ref{fig:ablation_more_translations1} shows all-pairs single-source translation from a single \ourmethod\ model: for each subject, every contrast is translated to every other contrast. Table~\ref{tab:multi_source_translation_full} additionally reports multi-source translation, where one, two, or three input contrasts jointly condition a single target. Together, they show that the same model translates between arbitrary contrast pairs and improves as more sources are provided.

\begin{figure*}[htbp]
\begin{center}
\includegraphics[width=1.0\textwidth]{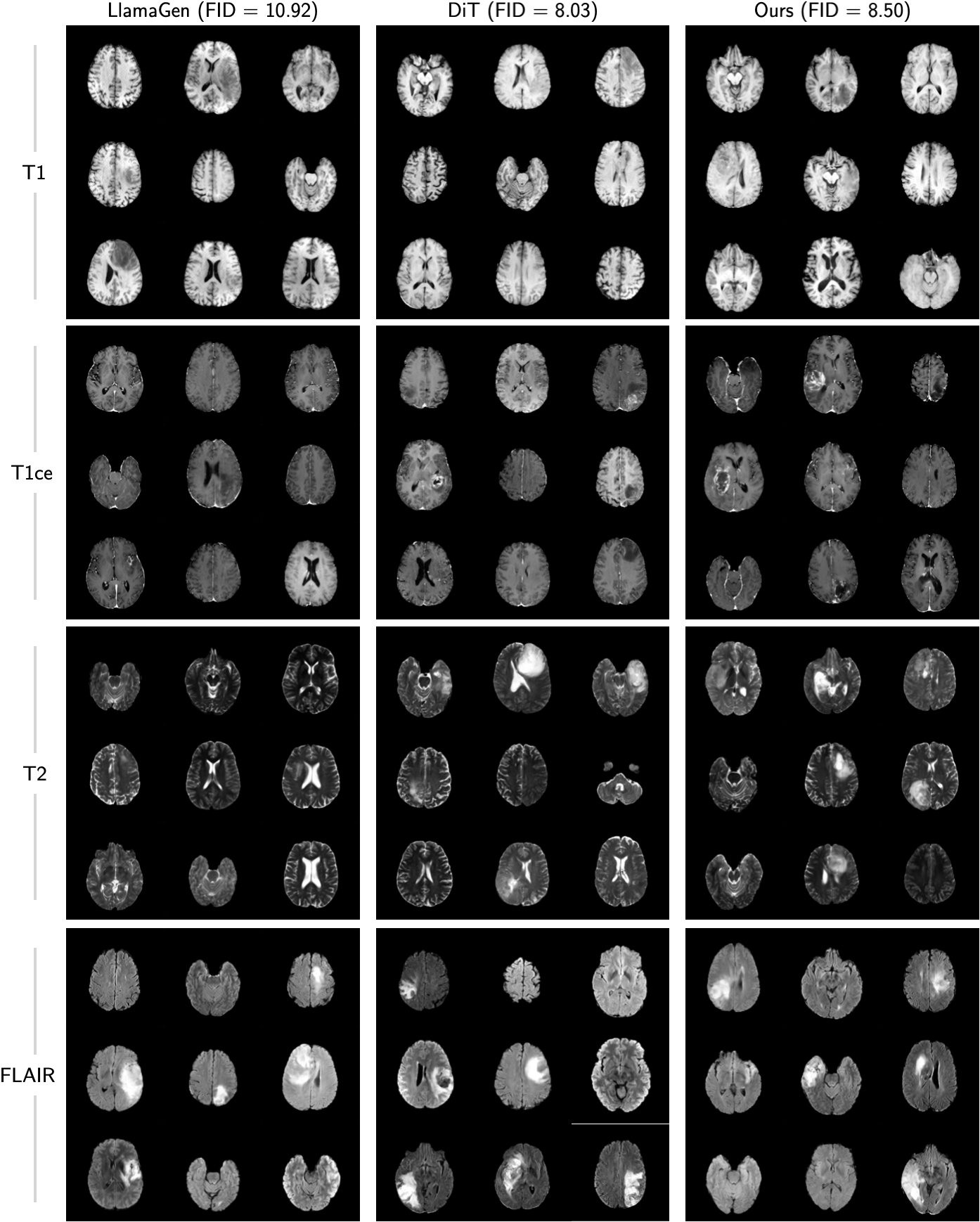}
\end{center}
\caption{\textbf{Qualitative comparison across methods on BraTS dataset.} Generated multi-contrast MRI samples from LlamaGen~\cite{sun2024Llamagen} (generation-only, raster autoregressive), DiT~\cite{peebles2023dit} (generation-only, latent diffusion), and \ourmethod\ (unified, next-dense-stride autoregressive).}
\label{fig:ablation_different_method_on_brats}
\vspace{-.3cm}
\end{figure*}

\paragraph{Modality-conditioned MRI generation.} Figure~\ref{fig:ablation_different_method_on_brats} compares generated samples from LlamaGen~\cite{sun2024Llamagen}, DiT~\cite{peebles2023dit}, and \ourmethod\ on BraTS~\cite{Menzeetal2015brats1, Bakasetal2018brats2}. LlamaGen and DiT are class-conditioned generation-only models, one autoregressive and one diffusion-based, while \ourmethod\ produces comparable samples from a single backbone that also performs translation and segmentation.

\begin{figure*}[htbp]
\begin{center}
\includegraphics[width=1.0\textwidth]{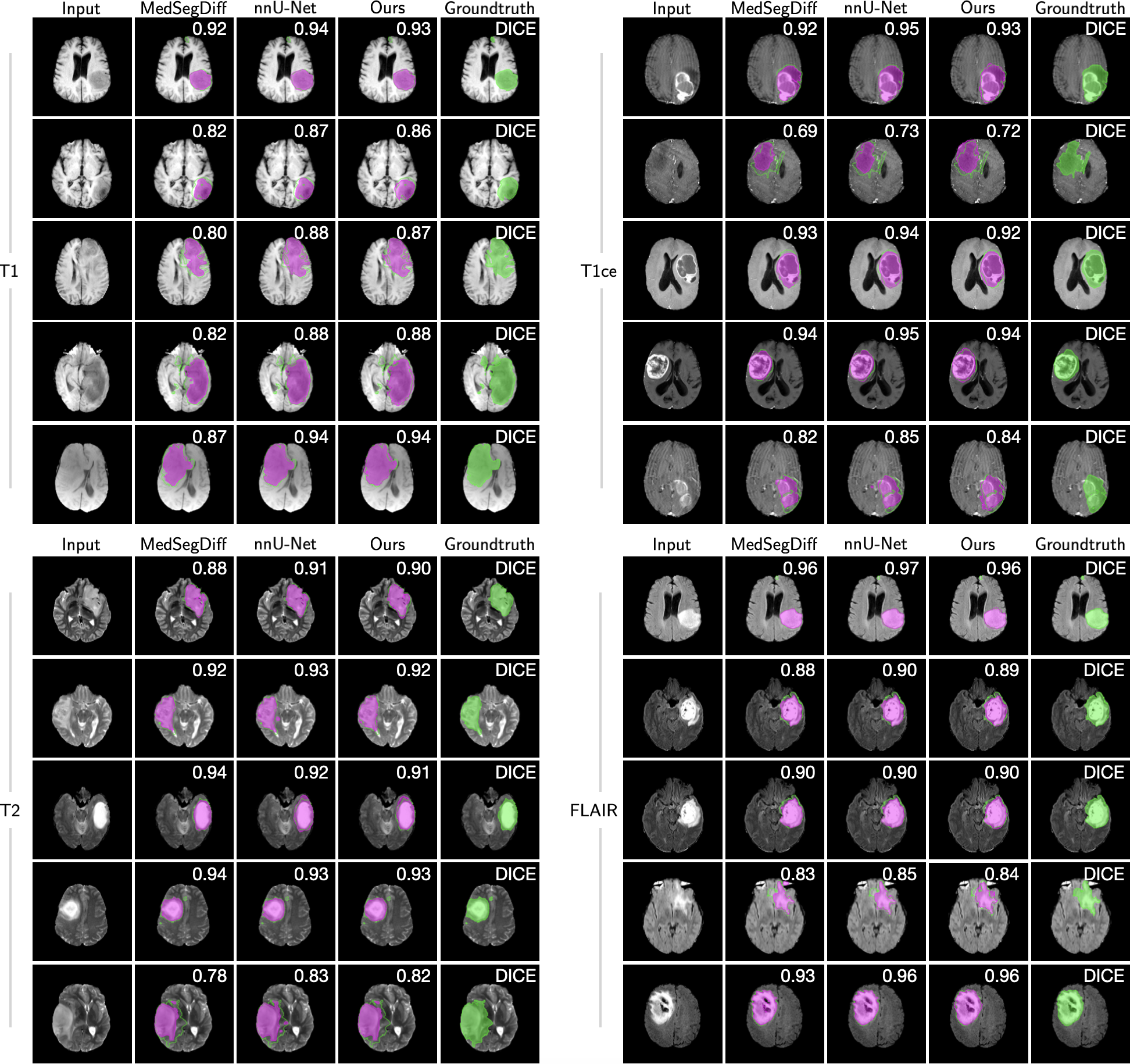}
\end{center}
\caption{\textbf{Tumor segmentation across methods and input modalities on BraTS-2023.} Each block group's examples are shown by the input contrast (T1, T1ce, T2, FLAIR); columns show the input, MedSegDiff-V2~\cite{wu2023medsegdiffv2} , nnU-Net~\cite{Isenseeetal2021nnUNet}, \ourmethod, and the ground truth, with Dice reported per example.
The \textcolor{magenta}{pink mask} is the predicted segmentation and the \textcolor{green}{green mask} is the ground truth. \ourmethod\ produces masks comparable to the task-specific baselines using the same token interface it uses for generation and translation, with no segmentation-specific head.}
\label{fig:ablation_more_segmentation}
\vspace{-.3cm}
\end{figure*}

\paragraph{Tumor segmentation.} Figure~\ref{fig:ablation_more_segmentation} compares tumor segmentation from MedSegDiff, nnU-Net, and \ourmethod\ across different input contrasts. \ourmethod\ produces masks comparable to the task-specific baselines through the same token interface it uses for generation and translation, with no segmentation-specific head.

\subsection{Natural Image Experiment}

\begin{figure*}[htbp]
\begin{center}
\includegraphics[width=1.0\textwidth]{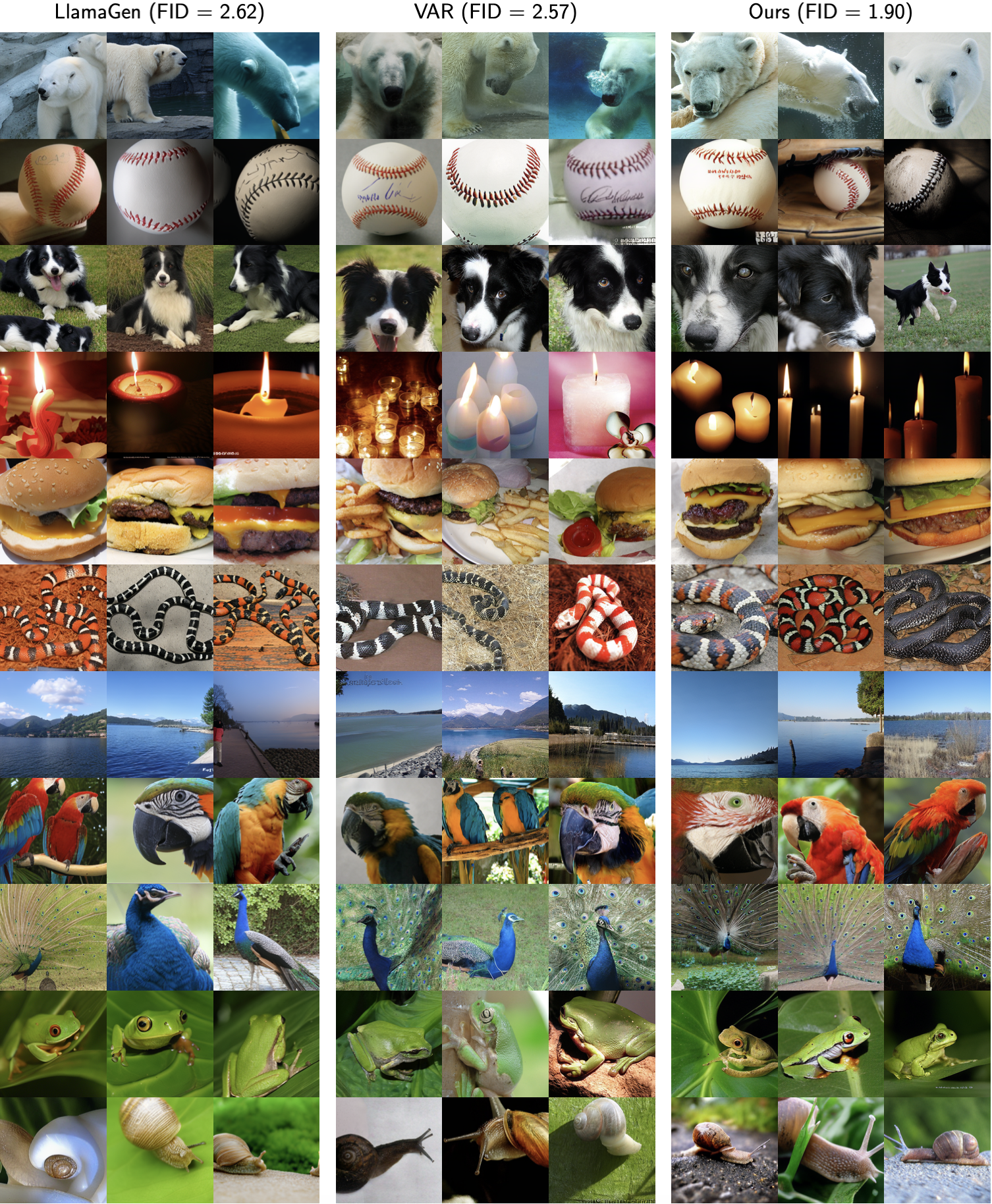}
\end{center}
\caption{\textbf{Qualitative comparison across methods on ImageNet.} Class-conditional samples from LlamaGen-XL~\cite{llamagen} (raster order), VAR-d20~\cite{var} (next-scale order), and \ourmethod-XL (next-dense-stride order).}
\vspace{-.3cm}
\label{fig:ablation_different_method}
\end{figure*}

\begin{figure*}[htbp]
\begin{center}
\includegraphics[width=0.95\textwidth]{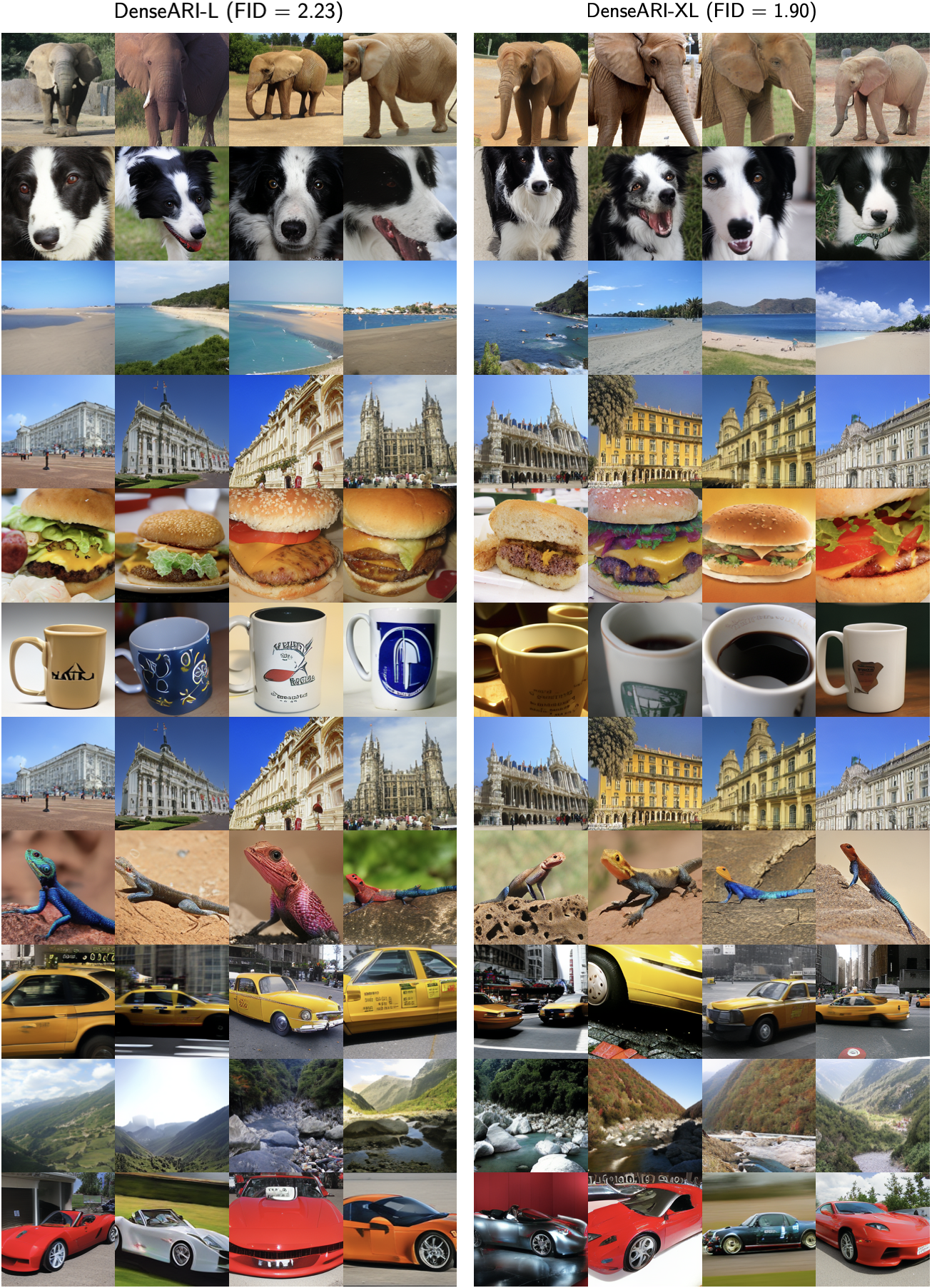}
\end{center}
\caption{\textbf{Effect of model scale.} Class-conditional samples from \ourmethod-L and -XL under identical class labels.}
\vspace{-.3cm}
\label{fig:ablation_different_weight}
\end{figure*}

Figure~\ref{fig:ablation_different_method} compares class-conditional generated samples from LlamaGen~\cite{sun2024Llamagen}, VAR~\cite{var}, and \ourmethod\ on ImageNet~\cite{deng2009imagenet}. We select these two baselines because they embody the limitations \ourmethod\ addresses: raster-order slowness and multi-scale sequence inflation, respectively. Figure~\ref{fig:ablation_different_weight} shows how sample quality scales with model size, comparing \ourmethod-L and -XL.

\subsection{Effect of Task-conditional Classifier-free Guidance}
\label{appendix:cfg-effect}

Section~\ref{sec:method} introduces a task-conditional form of classifier-free guidance, where the guidance weight $w$ controls how strongly each prediction follows its source modalities. Because the three tasks demand different degrees of source adherence, we do not fix a single $w$; instead, we sweep $w$ from $0.5$ to $15$ for each task and select the value that optimizes that task's metric (Figure~\ref{fig:ablation_guidance}).

The three tasks clearly favor different guidance strengths. Translation improves monotonically with $w$ and peaks at the largest setting ($w{=}15$, PSNR $21.9$): reconstructing a target contrast from a source scan benefits from strong adherence to the source. Generation prefers the opposite regime, reaching its best RadFID at a small weight ($w{=}4$) and degrading as $w$ grows, since sampling from a modality prior needs diversity rather than tight source conditioning. Segmentation sits between the two, peaking at $w{=}6$. Each task also shows a plateau (shaded) around its optimum, so the exact choice of $w$ is not brittle, but the best regions do not overlap across tasks.

These curves support the design in Section~\ref{sec:method}: a single shared guidance strength would compromise at least one task, whereas setting $w$ per task lets one model serve all three without retraining. The per-task values used in the main results are listed in Section~\ref{sec:numericalevaluations}.

\section{Implementation Details}
\label{appendix:implementation_details}

\subsection{More Visual Illustration of \ourmethod}
\label{appendix:visual_illustration_of_ours}

\begin{figure*}[htbp]
\begin{center}
\includegraphics[width=0.7\textwidth]{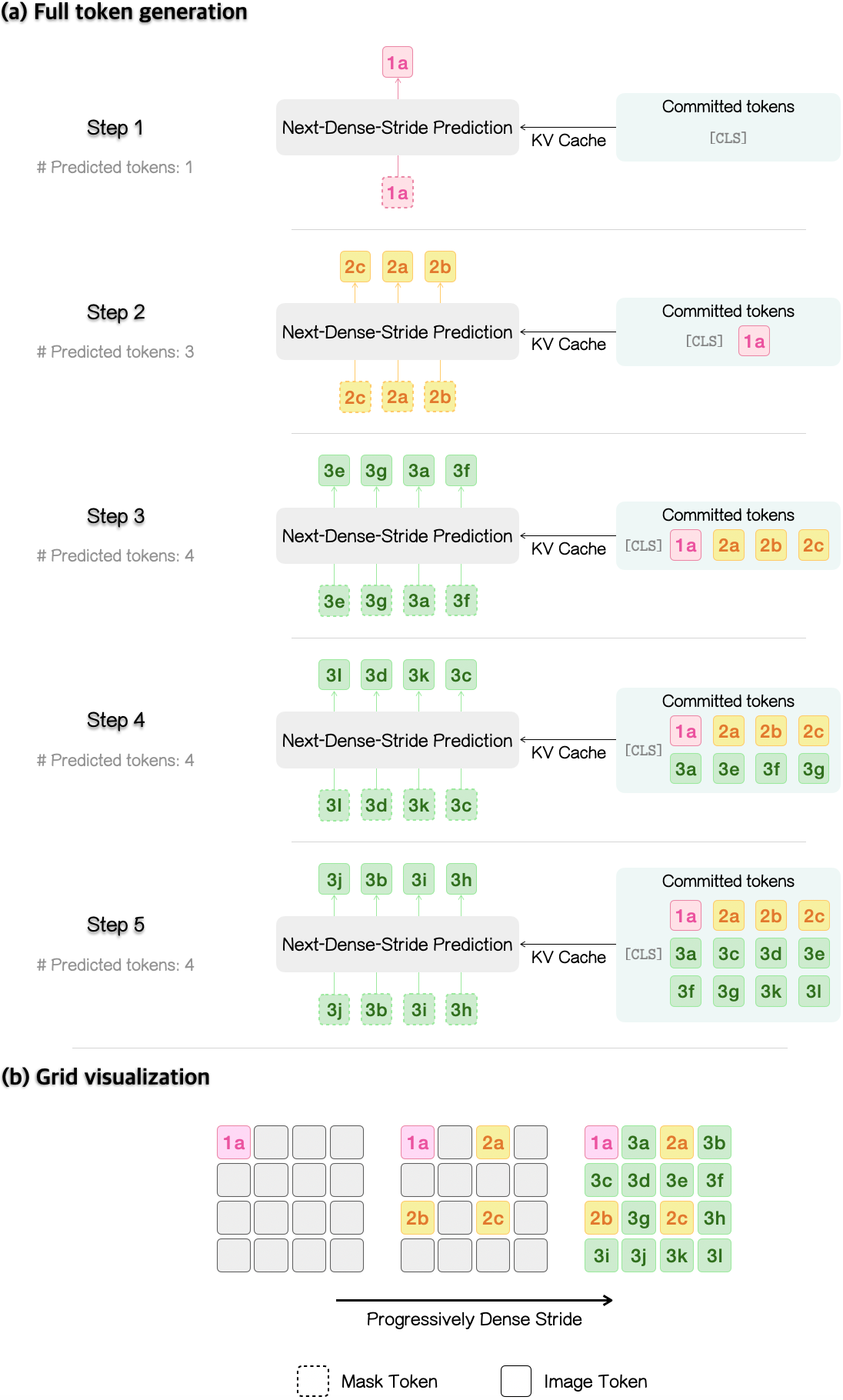}
\end{center}
\caption{\textbf{Step-by-step inference of \ourmethod\ (illustration only).} Here a $4\times4$ token grid ($16$ tokens) is generated over four steps, predicting $3$, $4$, $4$, and $5$ tokens per step. The number of steps and the tokens predicted at each are fixed by the inference schedule; these particular values and the $4\times4$ grid are illustrative, not the sizes used in practice. \textbf{(a) Full token generation.} At each step, the model predicts the next group of tokens in parallel from their position-embedded mask-token queries, conditioned on the committed tokens in the key--value (KV) cache. Starting from \texttt{[CLS]}, each predicted group is appended to the committed set and reused as context, so tokens are encoded once rather than recomputed. Colors denote the stride stage (stage~1 \textcolor{pink}{pink}, stage~2 \textcolor{YellowOrange}{yellow}, stage~3 \textcolor{Green}{green}). \textbf{(b) Grid visualization.} The same tokens on the grid show generation proceeding from a few widely spaced positions to progressively denser strides, coarse to fine.}
\vspace{-.3cm}
\label{fig:ablation_generation_steps}
\end{figure*}

Figure~\ref{fig:ablation_generation_steps} illustrates how \ourmethod\ samples tokens step by step.
Sparse strides are predicted first to lay down the global structure, and the denser later strides, which contain more tokens, are predicted in parallel within each step.

\subsection{Hyperparameters for Baselines}
\label{appendix:baseline_config}
Table~\ref{tab:baseline_config} lists the training configuration for \ourmethod\ and every baseline, grouped by task.
We take each method from its official implementation and keep the suggested settings unless noted.

EditAR conditions on a text instruction. We use the following input prompts to condition the task.
\begin{itemize}[leftmargin=*]
  \item Translation, single source: ``Given the \{Source\} image, generate the corresponding \{Target\} image.''
  \item Translation, multiple sources: ``Given the \{Source 1\}, \{Source 2\}, and \{Source 3\} images, generate the corresponding \{Target\} image.''
  \item Generation: ``Generate a brain MRI \{Target\} slice.''
\end{itemize}

\begin{table}[htbp]
\centering
\caption{Training configurations for all baselines and \ourmethod, grouped by task. LR represents the initial learning rate and schedule is learning rate schedule.}
\label{tab:baseline_config}
\resizebox{\linewidth}{!}{%
\begin{tabular}{llcccccl}
\toprule
Task & Method & \#Params & Epochs & Batch & LR & Schedule \\
\midrule
\multirow{3}{*}{Translation}
  & Pix2Pix~\cite{pix2pix2017}   & 57M  & 100 & 16 & $2{\times}10^{-4}$ & linear   \\
  & CycleGAN~\cite{CycleGAN2017} & 28M  & 100 & 16 & $8{\times}10^{-4}$ & linear   \\
  & MSG-LDM~\cite{lin2026msgldm} & 139M & 100 & 9  & $1{\times}10^{-4}$ & constant \\
\midrule
\multirow{2}{*}{Generation}
  & LlamaGen-L~\cite{llamagen} & 340M & 100 & 128 & $1{\times}10^{-4}$ & constant \\
  & DiT~\cite{peebles2023dit}  & 305M & 100 & 128 & $1{\times}10^{-4}$ & constant \\
\midrule
\multirow{2}{*}{Segmentation}
  & nnU-Net~\cite{Isenseeetal2021nnUNet} & 93M  & 100 & 64 & $1{\times}10^{-2}$ & poly   \\
  & MedSegDiff~\cite{wu2023medsegdiffv2} & 129M & 100 & 32 & $1{\times}10^{-4}$ & linear \\
\midrule
\multirow{3}{*}{Generalist}
  & EditAR~\cite{mu2025editAR} & 775M & 100 & 64  & $1{\times}10^{-4}$ & constant \\
  & MVG~\cite{ren2024MVG}      & 371M & 100 & 128 & $1{\times}10^{-3}$ & cosine   \\
  & Ours                       & 719M & 100 & 128 & $1{\times}10^{-3}$ & cosine   \\
\bottomrule
\end{tabular}%
}
\end{table}

\section{Discussions}
\label{appendix:more_discussions}

\paragraph{Generalist versus specialist.}
Medical imaging pipelines are usually built from many narrow models---one for each contrast-to-contrast translation, one for generation, one for each segmentation target---and each must be separately trained with different objectives, validated, and maintained.
As the number of modalities and tasks grows, this collection becomes costly to operate: adding a new modality or task means building and integrating yet another model. A single backbone that handles all of these would remove much of this overhead, but only if it does not unify tasks at the price of accuracy.

\ourmethod\ is such a backbone. Trained under one objective, it covers cross-modal translation, modality-conditioned generation, and segmentation, and because every modality reduces to the same token grid, a new modality or task can be absorbed by fine-tuning on its tokens rather than by adding a task-specific module, encoder, or head. Crucially, this breadth does not come at a large accuracy cost. On our medical benchmark, a single \ourmethod\ leads on translation, remains competitive on generation, and ranks a close second on segmentation, where a dedicated model such as nnU-Net still holds a modest edge. We do not claim to beat every specialist on its own task; rather, \ourmethod\ delivers all three from one model while staying close to the best specialist on each---turning a suite of narrow models into a single, extensible one.

\paragraph{Comparison to in-context medical generalists.}
The closest prior unified medical model is MVG~\cite{ren2024MVG}, which also casts several imaging tasks as image-to-image generation but specifies a task through in-context learning. MVG defines each task with an example pair: a source image and its ground-truth target, such as a scan and its segmentation mask, or a source-modality scan and the matching target-modality scan. This pair is stitched together with the test image onto a single canvas, and a masked vision transformer~\cite{dosovitskiy2020vit} fills in the held-out target under a regression loss. Two limitations follow. First, the example pair is the only task specification, so every prediction needs a fully labeled pair at inference rather than a lightweight task marker; for translation, this pair must already contain the target modality the model is meant to produce. Second, the single canvas holds one source image, so MVG cannot condition on several inputs at once, as multi-contrast synthesis requires.

\ourmethod\ instead represents each modality as a grid of discrete tokens and poses a task purely as a choice of which grids are given and which are predicted, marked by a short token in the sequence. This removes both limitations: a task is specified by a marker rather than a labeled example pair, and the sequence can hold zero, one, or several source grids, so \ourmethod\ generates from a modality marker with no source, translates from a single source, or combines several sources for multi-contrast synthesis, and produces both continuous images and discrete segmentation maps through the same interface.

\paragraph{Limitations and future work.}
\ourmethod\ is demonstrated only on visual tasks; it has not yet been shown to generate two modalities at once, such as image and text. Because \ourmethod\ is an autoregressive transformer operating on discrete tokens---the same paradigm used by language models---the sequence could in principle hold text tokens alongside image tokens, letting one model condition on or emit text within the same next-dense-stride pipeline. We do not evaluate this here and present it only as a promising direction rather than a contribution.

\section*{Disclaimer}

The views, methods, and results presented in this work are those of the authors and do not necessarily represent the official views, products, or commercial positions of GE HealthCare.

\begin{table*}[t]
\centering
\caption{Quantitative comparison of \textbf{multi-source} multimodal translation on BraTS 2023 across all target modalities. 
The input consists of 1, 2, or 3 source modalities to synthesize a single target modality. 
$\uparrow$/$\downarrow$: higher/lower is better. \hlgreen{Best} and \hlblue{second-best} results are color-coded per row configuration. Note that MVG~\cite{ren2024MVG} is not included because it cannot handle multi-source translation.}
\label{tab:multi_source_translation_full}
\renewcommand{\arraystretch}{1.2}
\setlength{\tabcolsep}{3.5pt}
\resizebox{\textwidth}{!}{%
\begin{tabular}{clc cccc cccc cccc}
\toprule
\multirow{2}{*}{\textbf{\# Src}} & \multirow{2}{*}{\textbf{Source Modalities}} & \multirow{2}{*}{\textbf{Target}}
& \multicolumn{4}{c}{EditAR~\cite{mu2025editAR}}
& \multicolumn{4}{c}{MSG-LDM~\cite{lin2026msgldm}}
& \multicolumn{4}{c}{\textbf{Ours}} \\
\cmidrule(lr){4-7} \cmidrule(lr){8-11} \cmidrule(lr){12-15}
& & & PSNR$\uparrow$ & SSIM$\uparrow$ & LPIPS$\downarrow$ & L1$\downarrow$
& PSNR$\uparrow$ & SSIM$\uparrow$ & LPIPS$\downarrow$ & L1$\downarrow$
& PSNR$\uparrow$ & SSIM$\uparrow$ & LPIPS$\downarrow$ & L1$\downarrow$ \\
\midrule

\multirow{3}{*}{1} 
  & T1ce  & \multirow{7}{*}{\textbf{T1}} & 21.38 & 0.837 & 0.106 & 0.036 & \hlblue{22.69} & \hlblue{0.862} & \hlblue{0.096} & \hlblue{0.031} & \hlgreen{23.19} & \hlgreen{0.874} & \hlgreen{0.086} & \hlgreen{0.029} \\
& T2     & & 20.69 & 0.833 & 0.115 & 0.039 & \hlblue{21.45} & \hlblue{0.850} & \hlblue{0.108} & \hlblue{0.036} & \hlgreen{22.50} & \hlgreen{0.871} & \hlgreen{0.092} & \hlgreen{0.031} \\
& FLAIR  & & 20.36 & 0.813 & 0.116 & 0.041 & \hlblue{21.32} & \hlblue{0.832} & \hlblue{0.110} & \hlblue{0.036} & \hlgreen{22.32} & \hlgreen{0.857} & \hlgreen{0.092} & \hlgreen{0.032} \\
\cmidrule(lr){1-2}\cmidrule(lr){4-15}
\multirow{3}{*}{2} 
  & T1ce + T2    & & 21.97 & 0.853 & 0.100 & 0.034 & \hlgreen{23.19} & \hlgreen{0.875} & \hlblue{0.089} & \hlgreen{0.029} & \hlblue{23.01} & \hlblue{0.871} & \hlgreen{0.085} & \hlblue{0.029} \\
& T1ce + FLAIR   & & 21.86 & 0.846 & 0.099 & 0.034 & \hlgreen{23.17} & \hlgreen{0.870} & \hlblue{0.089} & \hlgreen{0.030} & \hlblue{23.23} & \hlblue{0.873} & \hlgreen{0.083} & \hlgreen{0.029} \\
& T2 + FLAIR     & & 21.29 & 0.843 & 0.106 & 0.036 & \hlblue{22.20} & \hlblue{0.860} & \hlblue{0.098} & \hlblue{0.033} & \hlgreen{22.77} & \hlgreen{0.871} & \hlgreen{0.087} & \hlgreen{0.030} \\
\cmidrule(lr){1-2}\cmidrule(lr){4-15}
3 & T1ce + T2 + FLAIR & & 22.15 & 0.855 & 0.097 & 0.033 & \hlgreen{23.49} & \hlgreen{0.879} & \hlblue{0.085} & \hlgreen{0.029} & \hlblue{23.37} & \hlblue{0.877} & \hlgreen{0.081} & \hlblue{0.028} \\

\midrule
\midrule

\multirow{3}{*}{1} 
  & T1    & \multirow{7}{*}{\textbf{T1ce}} & 21.14 & 0.825 & 0.115 & 0.038 & \hlblue{21.57} & \hlblue{0.832} & \hlblue{0.111} & \hlblue{0.036} & \hlgreen{22.42} & \hlgreen{0.856} & \hlgreen{0.097} & \hlgreen{0.032} \\
& T2     & & 20.50 & 0.812 & 0.125 & 0.042 & \hlblue{21.07} & \hlblue{0.823} & \hlblue{0.117} & \hlblue{0.038} & \hlgreen{21.93} & \hlgreen{0.846} & \hlgreen{0.105} & \hlgreen{0.034} \\
& FLAIR  & & 20.28 & 0.799 & 0.127 & 0.043 & \hlblue{21.07} & \hlblue{0.813} & \hlblue{0.118} & \hlblue{0.037} & \hlgreen{21.72} & \hlgreen{0.836} & \hlgreen{0.106} & \hlgreen{0.034} \\
\cmidrule(lr){1-2}\cmidrule(lr){4-15}
\multirow{3}{*}{2} 
  & T1 + T2      & & 21.38 & 0.830 & 0.111 & 0.037 & \hlblue{22.02} & \hlblue{0.841} & \hlblue{0.106} & \hlblue{0.034} & \hlgreen{22.40} & \hlgreen{0.851} & \hlgreen{0.097} & \hlgreen{0.032} \\
& T1 + FLAIR     & & 21.37 & 0.828 & 0.111 & 0.037 & \hlblue{21.99} & \hlblue{0.838} & \hlblue{0.106} & \hlgreen{0.033} & \hlgreen{22.41} & \hlgreen{0.853} & \hlgreen{0.097} & \hlgreen{0.032} \\
& T2 + FLAIR     & & 20.88 & 0.818 & 0.118 & 0.039 & \hlblue{21.68} & \hlblue{0.831} & \hlblue{0.110} & \hlblue{0.035} & \hlgreen{22.10} & \hlgreen{0.848} & \hlgreen{0.101} & \hlgreen{0.033} \\
\cmidrule(lr){1-2}\cmidrule(lr){4-15}
3 & T1 + T2 + FLAIR & & 21.46 & 0.831 & 0.110 & 0.037 & \hlblue{22.29} & \hlblue{0.844} & \hlblue{0.103} & \hlgreen{0.032} & \hlgreen{22.38} & \hlgreen{0.847} & \hlgreen{0.098} & \hlgreen{0.032} \\

\midrule
\midrule

\multirow{3}{*}{1} 
  & T1    & \multirow{7}{*}{\textbf{T2}} & 21.02 & 0.827 & 0.110 & 0.038 & \hlblue{21.68} & \hlblue{0.842} & \hlblue{0.104} & \hlblue{0.035} & \hlgreen{22.54} & \hlblue{0.863} & \hlgreen{0.090} & \hlgreen{0.032} \\
& T1ce   & & 20.80 & 0.818 & 0.112 & 0.038 & \hlblue{21.75} & \hlblue{0.839} & \hlblue{0.103} & \hlblue{0.035} & \hlblue{22.38} & \hlblue{0.855} & \hlgreen{0.092} & \hlgreen{0.032} \\
& FLAIR  & & 20.41 & 0.804 & 0.116 & 0.041 & \hlblue{21.52} & \hlblue{0.829} & \hlblue{0.106} & \hlblue{0.035} & \hlgreen{22.33} & \hlblue{0.851} & \hlgreen{0.092} & \hlgreen{0.032} \\
\cmidrule(lr){1-2}\cmidrule(lr){4-15}
\multirow{3}{*}{2} 
  & T1 + T1ce    & & 21.42 & 0.835 & 0.104 & 0.036 & \hlblue{22.32} & \hlblue{0.854} & \hlblue{0.096} & \hlblue{0.033} & \hlgreen{22.88} & \hlgreen{0.868} & \hlgreen{0.087} & \hlgreen{0.031} \\
& T1 + FLAIR     & & 21.59 & 0.838 & 0.102 & 0.035 & \hlblue{22.49} & \hlgreen{0.856} & \hlblue{0.096} & \hlgreen{0.032} & \hlgreen{22.79} & \hlgreen{0.864} & \hlgreen{0.087} & \hlgreen{0.031} \\
& T1ce + FLAIR   & & 21.39 & 0.830 & 0.104 & 0.036 & \hlgreen{22.56} & \hlgreen{0.854} & \hlblue{0.095} & \hlgreen{0.032} & \hlblue{22.78} & \hlblue{0.859} & \hlgreen{0.088} & \hlgreen{0.031} \\
\cmidrule(lr){1-2}\cmidrule(lr){4-15}
3 & T1 + T1ce + FLAIR & & 21.73 & 0.841 & 0.101 & 0.034 & \hlgreen{22.94} & \hlgreen{0.864} & \hlblue{0.091} & \hlgreen{0.030} & \hlblue{23.08} & \hlblue{0.870} & \hlgreen{0.085} & \hlblue{0.030} \\

\midrule
\midrule

\multirow{3}{*}{1} 
  & T1    & \multirow{7}{*}{\textbf{FLAIR}} & 20.53 & 0.806 & 0.117 & 0.042 & \hlblue{21.10} & \hlblue{0.814} & \hlblue{0.112} & \hlblue{0.039} & \hlgreen{21.91} & \hlgreen{0.839} & \hlgreen{0.096} & \hlgreen{0.034} \\
& T1ce   & & 20.30 & 0.800 & 0.120 & 0.043 & \hlblue{21.12} & \hlblue{0.813} & \hlblue{0.113} & \hlblue{0.039} & \hlgreen{21.73} & \hlgreen{0.834} & \hlgreen{0.100} & \hlgreen{0.035} \\
& T2     & & 20.41 & 0.800 & 0.120 & 0.043 & \hlblue{21.29} & \hlblue{0.819} & \hlblue{0.113} & \hlblue{0.039} & \hlgreen{22.20} & \hlblue{0.842} & \hlgreen{0.097} & \hlgreen{0.033} \\
\cmidrule(lr){1-2}\cmidrule(lr){4-15}
\multirow{3}{*}{2} 
  & T1 + T1ce    & & 20.83 & 0.811 & 0.111 & 0.040 & \hlblue{21.59} & \hlblue{0.823} & \hlblue{0.104} & \hlblue{0.037} & \hlgreen{22.08} & \hlgreen{0.844} & \hlgreen{0.094} & \hlgreen{0.034} \\
& T1 + T2        & & 21.21 & 0.816 & 0.108 & 0.039 & \hlgreen{22.08} & \hlgreen{0.832} & \hlblue{0.102} & \hlblue{0.035} & \hlblue{22.10} & \hlgreen{0.840} & \hlgreen{0.093} & \hlgreen{0.033} \\
& T1ce + T2      & & 21.05 & 0.812 & 0.111 & 0.039 & \hlgreen{22.04} & \hlgreen{0.831} & \hlblue{0.103} & \hlblue{0.035} & \hlblue{22.03} & \hlblue{0.837} & \hlgreen{0.096} & \hlgreen{0.034} \\
\cmidrule(lr){1-2}\cmidrule(lr){4-15}
3 & T1 + T1ce + T2 & & 21.19 & 0.817 & 0.108 & 0.039 & \hlgreen{22.30} & \hlblue{0.836} & \hlblue{0.099} & \hlblue{0.034} & \hlblue{22.40} & \hlgreen{0.849} & \hlgreen{0.091} & \hlgreen{0.032} \\

\bottomrule
\end{tabular}%
}
\end{table*}

\end{document}